\documentclass{elsarticle}

\usepackage{xspace}
\usepackage{ifthen}
\usepackage{ifpdf} 

\makeatletter

\providecommand{\IfPackageLoaded}[2]{\@ifpackageloaded{#1}{#2}{}}
\providecommand{\IfPackageNotLoaded}[2]{\@ifpackageloaded{#1}{}{#2}}
\providecommand{\IfElsePackageLoaded}[3]{\@ifpackageloaded{#1}{#2}{#3}}
\newboolean{chapteravailable}%
\setboolean{chapteravailable}{false}%

\ifcsname chapter\endcsname
  \setboolean{chapteravailable}{true}%
\else
  \setboolean{chapteravailable}{false}%
\fi

\providecommand{\IfChapterDefined}[1]{\ifthenelse{\boolean{chapteravailable}}{#1}{}}%
\providecommand{\IfElseChapterDefined}[2]{\ifthenelse{\boolean{chapteravailable}}{#1}{#2}}%

\newcommand{\IfNotDraft}[1]{\ifx\@draft\@undefined #1 \fi}
\newcommand{\IfNotDraftElse}[2]{\ifx\@draft\@undefined #1 \else #2 \fi}
\newcommand{\IfDraft}[1]{\ifx\@draft\@undefined \else #1 \fi}
\@ifundefined{frontmatter}{%
   
}{}
\@ifundefined{mainmatter}{%
   \newif\if@mainmatter\@mainmattertrue
   
}{}
\@ifundefined{backmatter}{%
   
}{}

\newcommand{\LoadPackagesNow}{}
\newcommand{\LoadPackageLater}[1]{%
   \g@addto@macro{\LoadPackagesNow}{%
      \usepackage{#1}%
   }%
}

\makeatother

\usepackage[T1]{fontenc}
\usepackage[english]{babel}
\usepackage[utf8]{inputenc}
\usepackage{xargs}
\usepackage[pdftex,dvipsnames]{xcolor}

\usepackage[colorinlistoftodos,textsize=tiny]{todonotes}
\newcommandx{\unsure}[2][1=]{\todo[linecolor=red,backgroundcolor=red!25,bordercolor=red,#1]{#2}}

\usepackage[
   centertags, 
   sumlimits,  
   intlimits,  
   namelimits, 
   fleqn,     
]{amsmath} %

\usepackage{mathrsfs} 
\usepackage{dsfont}   
\usepackage{amssymb}

\usepackage{multirow}
\usepackage{array}
\usepackage{xfrac}
\usepackage{calc}
\usepackage{graphicx}
\graphicspath{{images/}}
\usepackage{epstopdf}
\usepackage{adjustbox}
\usepackage[Symbolsmallscale]{upgreek} 
\usepackage{fixmath}
\usepackage[inline]{enumitem}

\usepackage{framed}
\usepackage{multicol}
\usepackage{nomencl}
\makenomenclature
\renewcommand*\nompreamble{\begin{multicols}{2}}
\renewcommand*\nompostamble{\end{multicols}}

\newcommand\nomunit[1]{\def\nomentryend{\hfill$\left[#1\right]$}}

\renewcommand\nomgroup[1]{ 
    \def\makelabel##1{##1} 
    \ifx#1F\relax 
        \item
        \item[\textit{Time-dependent functions}] 
    \fi 
    \ifx#1C\relax
        \item
        \item[\textit{Constant parameters}] 
    \fi 
    \ifx#1I\relax 
        \item
        \item[\textit{Indicies}] 
    \fi 
    \let\makelabel\nomlabel 
}

\usepackage{listings}

\lstdefinestyle{C++Style}
{	language=C++,
	tabsize=4,
    showstringspaces=false,
    commentstyle=\color{green}\ttfamily,
    keywordstyle=\color{blue},
    basicstyle=\footnotesize,
    stringstyle=\color{red},
	frameround=fttt,
	frame=single,
	resetmargins=true
}

\lstdefinestyle{PythonStyle}
{	language=Python,
	commentstyle=\color{codegreen},
    keywordstyle=\color{magenta},
    numberstyle=\tiny\color{codegray},
    stringstyle=\color{codepurple},
    basicstyle=\footnotesize,
    breakatwhitespace=false,
    breaklines=true,
    captionpos=t,
    keepspaces=true,
    showspaces=false,
    showstringspaces=false,
    showtabs=false,
    tabsize=2,
    frameround=fttt,
    frame=single,
    resetmargins=true
}

\lstdefinestyle{JavaStyle}
{	language=Java,
    commentstyle=\color{green}\ttfamily,
    keywordstyle=\color{RoyalBlue},
    basicstyle=\footnotesize,
    rulecolor=\color{black},
    upquote=true,
    showstringspaces=false,
    breaklines=true,
    captionpos=t,
    frameround=fttt,
    frame=single,
    resetmargins=true
}

\usepackage[
	breaklinks,              
	hidelinks,
	linktocpage=true,        
   pdfcreator={LaTeX, hyperref, KOMA-Script}, 
   pdfdisplaydoctitle=true, 
   pdfpagelabels=true,           
]{hyperref}

\usepackage{glossaries}
\glsdisablehyper 

\IfPackageLoaded{hyperref}{%
	\usepackage[figure,table]{hypcap}
}

\providecommand*{\figrefname}{Figure~}
\newcommand*{\figref}[1]{%
  \hyperref[#1]{\figrefname{}}\ref{#1}%
}
\providecommand*{\tabrefname}{Table~}
\newcommand*{\tabref}[1]{%
  \hyperref[#1]{\tabrefname{}}\ref{#1}%
}
\providecommand*{\secrefname}{Section }
\newcommand*{\secref}[1]{%
  \hyperref[#1]{\secrefname{}}\ref{#1}%
}
\providecommand*{\chaprefname}{Chapter~}

\providecommand*{\exmprefname}{Example~}

\providecommand*{\satzrefname}{Satz~}

\newcommand*{\picref}[2]{%
  \hyperref[#1]{#2\,\faExternalLink}%
}
\providecommand*{\algorefname}{Algorithm~}

\providecommand*{\eqnrefname}{eqn.~}

\providecommand*{\Eqnrefname}{Eqn.~}
\newcommand*{\Eqnref}[1]{%
	\hyperref[#1]{\Eqnrefname{}}\eqref{#1}%
}

\usepackage{pdfpages} 
\usepackage{breakurl}
\usepackage{lscape}
\usepackage{longtable}
\usepackage{tabularx}

\usepackage[Symbolsmallscale]{upgreek} 

\usepackage{extarrows}

\usepackage[
	per=frac,
	fraction=nice,
	binary-units=true,
	loctolang={DE:ngerman,UK:english},
]{siunitx}

\usepackage{chemformula}

\usepackage{dirtree}

\usepackage{nameref}

\usepackage{bigints}

\usepackage{filecontents}

\usepackage[Algorithmus]{algorithm}
\usepackage{algpseudocode}

\algdef{SE}[DOWHILE]{Do}{doWhile}{\algorithmicdo}[1]{\algorithmicwhile\ #1}%

\usepackage{amsthm}
\usepackage{thmtools}
\theoremstyle{definition}
\newtheorem*{remark*}{Remark}

\usepackage{bm}

\usepackage{etoolbox}

\newcommand{\dd}[1]{{\, \rm d}#1}                     

\newcommand{\conj}[1][non]{\ensuremath{ \ifthenelse{\equal{#1}{non}}{\ast}{#1^\ast} } }                                            

\newcommand{\degree}[1][non]{\ensuremath{\ifthenelse{\equal{#1}{non}}{^\circ}{#1^\circ}}}

\newcommand{\derit}[1][non]{\ensuremath{\ifthenelse{\equal{#1}{non}}{\textrm d_{t}}}{\textrm d_{t\left({#1}\right)}}}    
\newcommand{\deribt}[1][non]{\ensuremath{\ifthenelse{\equal{#1}{non}}{\textrm d_{\bart}}}{\textrm d_{\bart\left({#1}\right)}}}    
\newcommand{\deriz}[1][non]{\ensuremath{\ifthenelse{\equal{#1}{non}}{\textrm d_{z}}}{\textrm d_{z\left({#1}\right)}}}    
\newcommand{\pderiz}[1][non]{\ensuremath{\ifthenelse{\equal{#1}{non}}{\partial_{z}}}{\partial_{z\left({#1}\right)}}}    
\newcommand{\pderir}[1][non]{\ensuremath{\ifthenelse{\equal{#1}{non}}{\partial_{r}}}{\partial_{r\left({#1}\right)}}}    
\newcommand{\pderith}[1][non]{\ensuremath{\ifthenelse{\equal{#1}{non}}{\partial_{\vartheta}}}{\partial_{\vartheta\left({#1}\right)}}}    
\newcommand{\pderithth}[1][non]{\ensuremath{\ifthenelse{\equal{#1}{non}}{\partial_{\vartheta\vartheta}}}{\partial_{\vartheta\vartheta\left({#1}\right)}}}
\newcommand{\pderit}[1][non]{\ensuremath{\ifthenelse{\equal{#1}{non}}{\partial_{t}}}{\partial_{t\left({#1}\right)}}}    
\newcommand{\pderitz}[1][non]{\ensuremath{\ifthenelse{\equal{#1}{non}}{\partial_{tz}}}{\partial_{t\left({#1}\right)z}}} 	
\newcommand{\pderibt}[1][non]{\ensuremath{\ifthenelse{\equal{#1}{non}}{\partial_{\bar{t}}}}{\partial_{\bar{t}\left({#1}\right)}}}    
\newcommand{\pderibtz}[1][non]{\ensuremath{\ifthenelse{\equal{#1}{non}}{\partial_{\bar{t}z}}}{\partial_{\bar{t}\left({#1}\right)z}}}
\newcommand{\pderiphi}[1][non]{\ensuremath{\ifthenelse{\equal{#1}{non}}{\partial_{\varphi}}}{\partial_{\varphi\left({#1}\right)}}}

\newcommand{\QA}[1][non]{\ensuremath{\mathbb{A}}}
\newcommand{\QAp}[1][non]{\ensuremath{\mathbb{A}_{\mathrm{p}}}}
\newcommand{\Am}[1][non]{\ensuremath{A_{\mathrm{m}}}}
\newcommand{\Um}[1][non]{\ensuremath{U_{\mathrm{m}}}}
\newcommand{\Rm}[1][non]{\ensuremath{R_{\mathrm{m}}}}
\newcommand{\barRm}[1][non]{\ensuremath{\bar{R}_{\mathrm{m}}}}
\newcommand{\QAm}[1][non]{\ensuremath{\mathbb{A}_{\mathrm{m}}}}
\newcommand{\Aw}[1][non]{\ensuremath{A_{\mathrm{w}}}}
\newcommand{\As}[1][non]{\ensuremath{A_{\mathrm{s}}}}
\newcommand{\Rw}[1][non]{\ensuremath{R_{\mathrm{w}}}}
\newcommand{\barRw}[1][non]{\ensuremath{\bar{R}_{\mathrm{w}}}}
\newcommand{\Uw}[1][non]{\ensuremath{U_{\mathrm{w}}}}
\newcommand{\QAw}[1][non]{\ensuremath{\mathbb{A}_{\mathrm{w}}}}
\newcommand{\uOmegap}[1][non]{\ensuremath{\ifthenelse{\equal{#1}{non}}{\upOmega_{\mathrm{p}}}{\upOmega_{p,\mathrm{p,#1}}}}}
\newcommand{\baruOmegap}[1][non]{\ensuremath{\ifthenelse{\equal{#1}{non}}{\bar{\upOmega}_{\mathrm{p}}}{\bar{\upOmega}_{\mathrm{p,#1}}}}}
\newcommand{\uOmegam}[1][non]{\ensuremath{\ifthenelse{\equal{#1}{non}}{\upOmega_{\mathrm{m}}}{\upOmega_{\mathrm{m,#1}}}}}
\newcommand{\uOmegaw}[1][non]{\ensuremath{\ifthenelse{\equal{#1}{non}}{\upOmega_{\mathrm{w}}}{\upOmega_{\mathrm{w,#1}}}}}
\newcommand{\Est}[1][non]{\ensuremath{\ifthenelse{\equal{#1}{non}}{E_{\mathrm{st}}}{E_{\mathrm{st,#1}}}}}
\newcommand{\Ein}[1][non]{\ensuremath{\ifthenelse{\equal{#1}{non}}{E_{\mathrm{in}}}{E_{\mathrm{in,#1}}}}}
\newcommand{\Eout}[1][non]{\ensuremath{\ifthenelse{\equal{#1}{non}}{E_{\mathrm{out}}}{E_{\mathrm{out,#1}}}}}
\newcommand{\Egen}[1][non]{\ensuremath{\ifthenelse{\equal{#1}{non}}{E_{\mathrm{gen}}}{E_{\mathrm{gen,#1}}}}}

\newcommand{\er}[1][non]{\ensuremath{\ifthenelse{\equal{#1}{non}}{\bm{e}_{r}}{\bm{e}_{r,#1}}}}
\newcommand{\ez}[1][non]{\ensuremath{\ifthenelse{\equal{#1}{non}}{\bm{e}_{z}}{\bm{e}_{z,#1}}}}
\renewcommand{\eth}[1][non]{\ensuremath{\ifthenelse{\equal{#1}{non}}{\bm{e}_{\vartheta}}{\bm{e}_{\vartheta\mathrm{#1}}}}}

\newcommand{\dz}[1][non]{\ensuremath{\ifthenelse{\equal{#1}{non}}{\Delta z}{\Delta z_{#1}}}}

\newcommand{\error}[1][non]{\ensuremath{\ifthenelse{\equal{#1}{non}}{\mathrm{E}}{\mathrm{E}_{#1}}}}

\renewcommand{\Re}{\ensuremath{\mathbb{R}}}

\newcommand{\FT}[1][non]{\ensuremath{\mathfrak F\ifthenelse{\equal{#1}{non}}{}{ \left\{ #1 \right\}}}}
\newcommand{\IFT}[1][non]{\ensuremath{\mathfrak F^{-1}\ifthenelse{\equal{#1}{non}}{}{ \left\{ #1 \right\} }} }

\newcommand{\Tm}[1][non]{\ensuremath{\ifthenelse{\equal{#1}{non}}{T_{\mathrm{m}}}{T_{\mathrm{m},#1}}}}
\newcommand{\bTm}[1][non]{\ensuremath{\ifthenelse{\equal{#1}{non}}{\mathbf{T}_{\mathrm{m}}}{\mathbf{T}_{\mathrm{m},#1}}}}
\newcommand{\bTmDel}[1][non]{\ensuremath{\ifthenelse{\equal{#1}{non}}{\mathbf{T}_{\mathrm{m}}^\mathrm{del}}{\mathbf{T}_{\mathrm{m},#1}^\mathrm{del}}}}
\newcommand{\TDel}[1][non]{\ensuremath{\ifthenelse{\equal{#1}{non}}{T_{\ast}^\mathrm{del}}{T_{\ast,#1}^\mathrm{del}}}}
\newcommand{\bT}[1][non]{\ensuremath{\ifthenelse{\equal{#1}{non}}{\mathbf{T}_{\ast}}{\mathbf{T}_{\ast,#1}}}}
\newcommand{\Tmin}[1][non]{\ensuremath{\ifthenelse{\equal{#1}{non}}{T_{\mathrm{m,in}}}{T_{\mathrm{m,in},#1}}}}
\newcommand{\Tmout}[1][non]{\ensuremath{\ifthenelse{\equal{#1}{non}}{T_{\mathrm{m,out}}}{T_{\mathrm{m,out},#1}}}}
\newcommand{\Tw}[1][non]{\ensuremath{\ifthenelse{\equal{#1}{non}}{T_{\mathrm{w}}}{T_{\mathrm{w},#1}}}}
\newcommand{\Ta}[1][non]{\ensuremath{\ifthenelse{\equal{#1}{non}}{T_{\infty}}{T_{\infty,#1}}}}
\newcommand{\bTa}[1][non]{\ensuremath{\ifthenelse{\equal{#1}{non}}{\mathbf{T}_{\infty}}{\mathbf{T}_{\infty,#1}}}}
\newcommand{\hatTm}[1][non]{\ensuremath{\ifthenelse{\equal{#1}{non}}{\widehat{T}_{\mathrm{m}}}{\widehat{T}_{\mathrm{m},#1}}}}
\newcommand{\hatTmin}{\ensuremath{\widehat{T}_{\mathrm{m,in}}}}

\newcommand{\hattildeTm}[1][non]{\ensuremath{\ifthenelse{\equal{#1}{non}}{\widehat{\tilde T}_{\mathrm{m}}}{\widehat{\tilde T}_{\mathrm{m},#1}}}}
\newcommand{\hatTw}[1][non]{\ensuremath{\ifthenelse{\equal{#1}{non}}{\widehat{T}_{\mathrm{w}}}{\widehat{T}_{\mathrm{w},#1}}}}
\newcommand{\hatTa}[1][non]{\ensuremath{\ifthenelse{\equal{#1}{non}}{\widehat{T}_{\infty}}{\widehat{T}_{\infty,#1}}}}
\newcommand{\barTm}[1][non]{\ensuremath{\ifthenelse{\equal{#1}{non}}{\bar{T}_{\mathrm{m}}}{\bar{T}_{\mathrm{m},#1}}}}
\newcommand{\bbarTm}[1][non]{\ensuremath{\ifthenelse{\equal{#1}{non}}{\bar{\mathbf{T}}_{\mathrm{m}}}{\bar{\mathbf{T}}_{\mathrm{m},#1}}}}
\newcommand{\bTmh}[1][non]{\ensuremath{\ifthenelse{\equal{#1}{non}}{\hat{\mathbf{T}}_{\mathrm{m}}}{\hat{\mathbf{T}}_{\mathrm{m},#1}}}}
\newcommand{\barTw}[1][non]{\ensuremath{\ifthenelse{\equal{#1}{non}}{\bar{T}_{\mathrm{w}}}{\bar{T}_{\mathrm{w},#1}}}}
\newcommand{\barTa}[1][non]{\ensuremath{\ifthenelse{\equal{#1}{non}}{\bar{T}_{\infty}}{\bar{T}_{\infty,#1}}}}
\newcommand{\barz}[1][non]{\ensuremath{\ifthenelse{\equal{#1}{non}}{\bar{z}}{\bar{z}_{#1}}}}
\newcommand{\bart}[1][non]{\ensuremath{\ifthenelse{\equal{#1}{non}}{\bar{t}}{\bar{t}_{#1}}}}
\newcommand{\tildeT}[1][non]{\ensuremath{\ifthenelse{\equal{#1}{non}}{\tilde{T}_{\ast}}{\tilde{T}_{\ast,#1}}}}
\newcommand{\tildeTw}[1][non]{\ensuremath{\ifthenelse{\equal{#1}{non}}{\tilde{T}_{\mathrm{w}}}{\tilde{T}_{\mathrm{w},#1}}}}
\newcommand{\tildeTa}[1][non]{\ensuremath{\ifthenelse{\equal{#1}{non}}{\tilde{T}_{\infty}}{\tilde{T}_{\infty,#1}}}}
\newcommand{\aTm}[1][non]{\ensuremath{\ifthenelse{\equal{#1}{non}}{\acute{T}_{\mathrm{m}}}{\acute{T}_{\mathrm{m},#1}}}}

\newcommand{\vm}[1][non]{\ensuremath{\ifthenelse{\equal{#1}{non}}{v_{\mathrm{m}}}{v_{\mathrm{m},#1}}}}
\newcommand{\dvm}[1][non]{\ensuremath{\ifthenelse{\equal{#1}{non}}{\dot{v}_{\mathrm{m}}}{\dot{v}_{\mathrm{m},#1}}}}
\newcommand{\bmv}[1][non]{\ensuremath{\ifthenelse{\equal{#1}{non}}{\mathbf{v}}{\mathbf{v}_{#1}}}}

\newcommand{\rhom}[1][non]{\ensuremath{\rho_{\mathrm{m}}}}
\newcommand{\rhow}[1][non]{\ensuremath{\ifthenelse{\equal{#1}{non}}{\rho_{\mathrm{w}}}{\rho_{\mathrm{w},#1}}}}

\newcommand{\cpm}[1][non]{\ensuremath{c_{\mathrm{p,m}}}}
\newcommand{\cpw}[1][non]{\ensuremath{c_{\mathrm{p,w}}}}

\newcommand{\lambdam}[1][non]{\ensuremath{\lambda_{\mathrm{m}}}}
\newcommand{\lambdaw}[1][non]{\ensuremath{\ifthenelse{\equal{#1}{non}}{\lambda_{\mathrm{w}}}{\lambda_{\mathrm{w},#1}}}}

\newcommand{\alphamw}[1][non]{\ensuremath{\ifthenelse{\equal{#1}{non}}{\alpha_{\mathrm{mw}}}{\alpha_{\mathrm{mw},#1}}}}
\newcommand{\balphamw}[1][non]{\ensuremath{\ifthenelse{\equal{#1}{non}}{\bar{\alpha}_{\mathrm{mw}}}{\bar{\alpha}_{\mathrm{mw},#1}}}}
\newcommand{\alphawa}[1][non]{\ensuremath{\ifthenelse{\equal{#1}{non}}{\alpha_{\mathrm{wa}}}{\alpha_{\mathrm{wa},#1}}}}
\newcommand{\balphawa}[1][non]{\ensuremath{\ifthenelse{\equal{#1}{non}}{\bar{\alpha}_{\mathrm{wa}}}{\bar{\alpha}_{\mathrm{wa},#1}}}}
\newcommand{\alphama}[1][non]{\ensuremath{\ifthenelse{\equal{#1}{non}}{\alpha_{\mathrm{ma}}}{\alpha_{\mathrm{ma},#1}}}}

\newcommand{\bmdq}[1][non]{\ensuremath{\ifthenelse{\equal{#1}{non}}{\dot{\bm{q}}}{\dot{\bm{q}}_{#1}}}}
\newcommand{\dq}[1][non]{\ensuremath{\ifthenelse{\equal{#1}{non}}{\dot{q}}{\dot{q}_{#1}}}}
\newcommand{\dQ}[1][non]{\ensuremath{\ifthenelse{\equal{#1}{non}}{\dot{Q}}{\dot{Q}_{#1}}}}

\newcommand{\mmed}[1][non]{\ensuremath{\ifthenelse{\equal{#1}{non}}{m_{\mathrm{m}}}{m_{\mathrm{m},#1}}}}

\newcommand{\bbm}[1][non]{\ensuremath{\ifthenelse{\equal{#1}{non}}{\mathbf{b}_\mathrm{m}}{\mathbf{b}_{\mathrm{m},#1}}}}
\newcommand{\bAm}[1][non]{\ensuremath{\ifthenelse{\equal{#1}{non}}{\mathbf{A}_\mathrm{m}}{\mathbf{A}_{\mathrm{m},#1}}}}
\newcommand{\bDm}[1][non]{\ensuremath{\ifthenelse{\equal{#1}{non}}{\mathbf{D}_\mathrm{m}}{\mathbf{D}_{\mathrm{m},#1}}}}

\newcommand{\besselI}{\bm{I}}
\newcommand{\T}[1][non]{\ensuremath{\ifthenelse{\equal{#1}{non}}{T_{\ast}}{T_{\ast,#1}}}}

\newcommand{\TmDel}[1][non]{\ensuremath{\ifthenelse{\equal{#1}{non}}{T_{\mathrm{m}}^{\mathrm{del}}}{T_{\mathrm{m},#1}^{\mathrm{del}}}}}
\newcommand{\TminDel}{\Tmin^{\text{del}}}

\newcommand{\hatTmDel}{\widehat{T}_{\textrm{m}}^{\text{del}}}
\newcommand{\hatTminDel}{\widehat{T}_{\textrm{m,in}}^{\text{del}}}
\newcommand{\TaDel}[1][non]{\ensuremath{\ifthenelse{\equal{#1}{non}}{T_{\infty}^{\mathrm{del}}}{T_{\infty,#1}^{\mathrm{del}}}}}
\newcommand{\TwDel}{\Tw^{\text{del}}}
\newcommand{\vmDel}{\vm^{\text{del}}}

\newcommand{\atan}[1][non]{\arctan\ensuremath{\ifthenelse{\equal{#1}{non}}{}{ \left( #1 \right)}}}
\newcommand{\E}[1][non]{\ensuremath{\ifthenelse{\equal{#1}{non}}{E}{E\left\{ #1 \right\}}}}
\newcommand{\chis}[2][non]{\ensuremath{\chi^2 \ifthenelse{\equal{#1}{non}}{}{ \left(#1,#2\right)}}}
\newcommand{\R}[2][xx]{\ensuremath{R_{#1}\ifthenelse{\equal{#2}{}}{}{\left( #2 \right)}}} 
\renewcommand{\S}[2][xx]{\ensuremath{S_{#1}\ifthenelse{\equal{#2}{}}{}{\left( #2 \right)}}} 
\newcommand{\Gammaf}[1][non]{\ensuremath{\Gamma\ifthenelse{\equal{#1}{non}}{}{ \left( #1 \right)}}}

\newbool{markChanges}
\newcommand{\markChange}[1]{%
    \ifbool{markChanges}{%
        \textcolor{blue!75}{#1}%
    }{%
        #1%
    }%
}

\usepackage{tikz}
\usepackage{circuitikz}
\usepackage{csvsimple}
\ctikzset{voltage/distance from node=.5}
\ctikzset{voltage/distance from line=.25}
\ctikzset{voltage/bump b/.initial=1.5}%

\usepackage{pgfplots}
\usetikzlibrary{plotmarks,math,positioning,shapes,arrows,backgrounds,circuits.logic.IEC,circuits.ee.IEC,decorations.pathmorphing,patterns,shapes.geometric,calc,fit,spy,matrix}

\makeatletter
\global\let\tikz@ensure@dollar@catcode=\relax

\pgfdeclarepatternformonly{north east lines wide}%
   {\pgfqpoint{-1pt}{-1pt}}%
   {\pgfqpoint{10pt}{10pt}}%
   {\pgfqpoint{9pt}{9pt}}%
   {
     \pgfsetlinewidth{0.4pt}
     \pgfpathmoveto{\pgfqpoint{0pt}{0pt}}
     \pgfpathlineto{\pgfqpoint{9.1pt}{9.1pt}}
     \pgfusepath{stroke}
    }

\pgfdeclarepatternformonly{north west lines wide}%
   {\pgfqpoint{-1pt}{-1pt}}%
   {\pgfqpoint{10pt}{10pt}}%
   {\pgfqpoint{9pt}{9pt}}%
   {
     \pgfsetlinewidth{0.4pt}
     \pgfpathmoveto{\pgfqpoint{0pt}{9.1pt}}
     \pgfpathlineto{\pgfqpoint{9.1pt}{0pt}}
     \pgfusepath{stroke}
    }

\pgfkeys{/pgf/.cd,
  parallelepiped offset x/.initial=2mm,
  parallelepiped offset y/.initial=2mm
}
\pgfdeclareshape{parallelepiped}
{
  \inheritsavedanchors[from=rectangle] 
  \inheritanchorborder[from=rectangle]
  \inheritanchor[from=rectangle]{north}
  \inheritanchor[from=rectangle]{north west}
  \inheritanchor[from=rectangle]{north east}
  \inheritanchor[from=rectangle]{center}
  \inheritanchor[from=rectangle]{west}
  \inheritanchor[from=rectangle]{east}
  \inheritanchor[from=rectangle]{mid}
  \inheritanchor[from=rectangle]{mid west}
  \inheritanchor[from=rectangle]{mid east}
  \inheritanchor[from=rectangle]{base}
  \inheritanchor[from=rectangle]{base west}
  \inheritanchor[from=rectangle]{base east}
  \inheritanchor[from=rectangle]{south}
  \inheritanchor[from=rectangle]{south west}
  \inheritanchor[from=rectangle]{south east}
  \backgroundpath{
    \southwest \pgf@xa=\pgf@x \pgf@ya=\pgf@y
    \northeast \pgf@xb=\pgf@x \pgf@yb=\pgf@y
    \pgfmathsetlength\pgfutil@tempdima{\pgfkeysvalueof{/pgf/parallelepiped offset x}}
    \pgfmathsetlength\pgfutil@tempdimb{\pgfkeysvalueof{/pgf/parallelepiped offset y}}
    \def\ppd@offset{\pgfpoint{\pgfutil@tempdima}{\pgfutil@tempdimb}}
    \pgfpathmoveto{\pgfqpoint{\pgf@xa}{\pgf@ya}}
    \pgfpathlineto{\pgfqpoint{\pgf@xb}{\pgf@ya}}
    \pgfpathlineto{\pgfqpoint{\pgf@xb}{\pgf@yb}}
    \pgfpathlineto{\pgfqpoint{\pgf@xa}{\pgf@yb}}
    \pgfpathclose
    \pgfpathmoveto{\pgfqpoint{\pgf@xb}{\pgf@ya}}
    \pgfpathlineto{\pgfpointadd{\pgfpoint{\pgf@xb}{\pgf@ya}}{\ppd@offset}}
    \pgfpathlineto{\pgfpointadd{\pgfpoint{\pgf@xb}{\pgf@yb}}{\ppd@offset}}
    \pgfpathlineto{\pgfpointadd{\pgfpoint{\pgf@xa}{\pgf@yb}}{\ppd@offset}}
    \pgfpathlineto{\pgfqpoint{\pgf@xa}{\pgf@yb}}
    \pgfpathmoveto{\pgfqpoint{\pgf@xb}{\pgf@yb}}
    \pgfpathlineto{\pgfpointadd{\pgfpoint{\pgf@xb}{\pgf@yb}}{\ppd@offset}}
  }
}

\pgfdeclareshape{zohswitch}{
  \inheritsavedanchors[from=rectangle]
  \inheritanchorborder[from=rectangle]
  \inheritanchor[from=rectangle]{center}
  \inheritanchor[from=rectangle]{base}
  \inheritanchor[from=rectangle]{north}
  \inheritanchor[from=rectangle]{north east}
  \inheritanchor[from=rectangle]{east}
  \inheritanchor[from=rectangle]{south east}
  \inheritanchor[from=rectangle]{south}
  \inheritanchor[from=rectangle]{south west}
  \inheritanchor[from=rectangle]{west}
  \inheritanchor[from=rectangle]{north west}
  \behindbackgroundpath{
    \southwest \pgf@xa=\pgf@x \pgf@ya=\pgf@y
    \northeast \pgf@xb=\pgf@x \pgf@yb=\pgf@y
    
    \pgfpathmoveto{\pgfpoint{\pgf@xa}{(\pgf@ya + \pgf@yb)/2}}
    \pgfpathlineto{\pgfpoint{\pgf@xb}{\pgf@yb}}
    
    \pgfsetlinewidth{.15mm}\pgfsetarrows{-}\pgfsetstrokecolor{black}\pgfusepath{stroke}
 }
}

\pgfdeclareshape{datastore}{
  \inheritsavedanchors[from=rectangle]
  \inheritanchorborder[from=rectangle]
  \inheritanchor[from=rectangle]{center}
  \inheritanchor[from=rectangle]{base}
  \inheritanchor[from=rectangle]{north}
  \inheritanchor[from=rectangle]{north east}
  \inheritanchor[from=rectangle]{east}
  \inheritanchor[from=rectangle]{south east}
  \inheritanchor[from=rectangle]{south}
  \inheritanchor[from=rectangle]{south west}
  \inheritanchor[from=rectangle]{west}
  \inheritanchor[from=rectangle]{north west}

  \backgroundpath{
    \southwest \pgf@xa=\pgf@x \pgf@ya=\pgf@y
    \northeast \pgf@xb=\pgf@x \pgf@yb=\pgf@y
    \pgfpathmoveto{\pgfpoint{\pgf@xa}{\pgf@ya}}
    \pgfpathlineto{\pgfpoint{\pgf@xb}{\pgf@ya}}
    \pgfpathmoveto{\pgfpoint{\pgf@xa}{\pgf@yb}}
    \pgfpathlineto{\pgfpoint{\pgf@xb}{\pgf@yb}}
 }
}

\pgfdeclaredecoration{coilup}{coil}
{
  \state{coil}[switch if less than=%
    1.5\pgfdecorationsegmentlength+%
    \pgfdecorationsegmentaspect\pgfdecorationsegmentamplitude+%
    \pgfdecorationsegmentaspect\pgfdecorationsegmentamplitude to last,
               width=+\pgfdecorationsegmentlength]
  {
    \pgfpathcurveto
    {\pgfpoint@oncoil{0    }{ 0.555}{1}}
    {\pgfpoint@oncoil{0.445}{ 1    }{2}}
    {\pgfpoint@oncoil{1    }{ 1    }{3}}
    \pgfpathmoveto{\pgfpoint@oncoil{1    }{-1    }{9}}
    \pgfpathcurveto
    {\pgfpoint@oncoil{0.445}{-1    }{10}}
    {\pgfpoint@oncoil{0    }{-0.555}{11}}
    {\pgfpoint@oncoil{0    }{ 0    }{12}}
  }
  \state{last}[width=.5\pgfdecorationsegmentlength+%
    \pgfdecorationsegmentaspect\pgfdecorationsegmentamplitude+%
    \pgfdecorationsegmentaspect\pgfdecorationsegmentamplitude,next state=final]
  {
    \pgfpathcurveto
    {\pgfpoint@oncoil{0    }{ 0.555}{1}}
    {\pgfpoint@oncoil{0.445}{ 1    }{2}}
    {\pgfpoint@oncoil{1    }{ 1    }{3}}
    \pgfpathmoveto{\pgfpoint@oncoil{2    }{ 0    }{6}}
  }
  \state{final}
  {
  \pgfpathmoveto{\pgfpointdecoratedpathlast}
  }
}

\pgfdeclaredecoration{coildown}{coil}
{
  \state{coil}[switch if less than=%
    1.5\pgfdecorationsegmentlength+%
    \pgfdecorationsegmentaspect\pgfdecorationsegmentamplitude+%
    \pgfdecorationsegmentaspect\pgfdecorationsegmentamplitude to last,
               width=+\pgfdecorationsegmentlength]
  {
    \pgfpathmoveto{\pgfpoint@oncoil{1    }{1    }{3}}
    \pgfpathcurveto
    {\pgfpoint@oncoil{1.555}{ 1    }{4}}
    {\pgfpoint@oncoil{2    }{ 0.555}{5}}
    {\pgfpoint@oncoil{2    }{ 0    }{6}}
    \pgfpathcurveto
    {\pgfpoint@oncoil{2    }{-0.555}{7}}
    {\pgfpoint@oncoil{1.555}{-1    }{8}}
    {\pgfpoint@oncoil{1    }{-1    }{9}}
  }
  \state{last}[width=.5\pgfdecorationsegmentlength+%
    \pgfdecorationsegmentaspect\pgfdecorationsegmentamplitude+%
    \pgfdecorationsegmentaspect\pgfdecorationsegmentamplitude,next state=final]
  {
    \pgfpathmoveto{\pgfpoint@oncoil{1    }{ 1    }{3}}
    \pgfpathcurveto
    {\pgfpoint@oncoil{1.555}{ 1    }{4}}
    {\pgfpoint@oncoil{2    }{ 0.555}{5}}
    {\pgfpoint@oncoil{2    }{ 0    }{6}}
  }
  \state{final}
  {
  \pgfpathlineto{\pgfpointdecoratedpathlast}
  }
}

\def\pgfpoint@oncoil#1#2#3{%
  \pgf@x=#1\pgfdecorationsegmentamplitude%
  \pgf@x=\pgfdecorationsegmentaspect\pgf@x%
  \pgf@y=#2\pgfdecorationsegmentamplitude%
  \pgf@xa=0.083333333333\pgfdecorationsegmentlength%
  \advance\pgf@x by#3\pgf@xa%
}

\pgfmathdeclarefunction{erf}{1}{%
  \begingroup
    \pgfmathparse{#1 > 0 ? 1 : -1}%
    \edef\sign{\pgfmathresult}%
    \pgfmathparse{abs(#1)}%
    \edef\x{\pgfmathresult}%
    \pgfmathparse{1/(1+0.3275911*\x)}%
    \edef\t{\pgfmathresult}%
    \pgfmathparse{%
      1 - (((((1.061405429*\t -1.453152027)*\t) + 1.421413741)*\t 
      -0.284496736)*\t + 0.254829592)*\t*exp(-(\x*\x))}%
    \edef\y{\pgfmathresult}%
    \pgfmathparse{(\sign)*\y}%
    \pgfmath@smuggleone\pgfmathresult%
  \endgroup
}

\newif\ifcuboidshade
\newif\ifcuboidemphedge

\tikzset{
  cuboid/.is family,
  cuboid,
  shiftx/.initial=0,
  shifty/.initial=0,
  dimx/.initial=3,
  dimy/.initial=3,
  dimz/.initial=3,
  scale/.initial=1,
  densityx/.initial=1,
  densityy/.initial=1,
  densityz/.initial=1,
  rotation/.initial=0,
  anglex/.initial=0,
  angley/.initial=90,
  anglez/.initial=225,
  scalex/.initial=1,
  scaley/.initial=1,
  scalez/.initial=0.5,
  front/.style={draw=black,fill=white},
  top/.style={draw=black,fill=white},
  right/.style={draw=black,fill=white},
  shade/.is if=cuboidshade,
  shadecolordark/.initial=black,
  shadecolorlight/.initial=white,
  shadeopacity/.initial=0.15,
  shadesamples/.initial=16,
  emphedge/.is if=cuboidemphedge,
  emphstyle/.style={thick},
}

\tikzset{
  path picture/.code=\tikz@addmode{\def\tikz@path@picture{#1\tikz@path@picture@extra}},
  path picture extra/.code={\def\tikz@path@picture@extra{#1}}
}
\let\tikz@path@picture@extra\pgfutil@empty

\makeatother

\tikzstyle{prozess} = [draw, thick, rounded corners, inner sep=.3cm]
\tikzstyle{function} = [draw, thick, circle]
\tikzstyle{ifunction} = [draw, thick, circle split,minimum size=30mm, font=\small]
\tikzstyle{datastore} = [draw, very thick, shape=datastore, inner sep=.3cm]
\tikzstyle{to} = [->, >=stealth', shorten >=1pt, semithick, font=\sffamily\footnotesize]
\tikzstyle{dto} = [->, dashed, >=stealth', shorten >=1pt, semithick, font=\sffamily\footnotesize]
\tikzstyle{wto} = [-, color=white, >=stealth', shorten >=1pt, semithick, font=\sffamily\footnotesize]
\tikzstyle{bto} = [<->, >=stealth', shorten >=1pt, semithick, font=\sffamily\footnotesize]
\tikzstyle{ccircle} = [path picture={ \draw[black] (path picture bounding box.south east) -- (path picture bounding box.north west) (path picture bounding box.south west) -- (path picture bounding box.north east);}]

\tikzstyle{block} = [draw, fill=white, rectangle, minimum height=3em, minimum width=4em]
\tikzstyle{rblock} = [draw, fill=white, circle, inner sep=0pt,minimum size=1mm]
\tikzstyle{wobblock} = [fill=white, rectangle, minimum height=3em, minimum width=5em]
\tikzstyle{nlblock} = [draw, postaction={draw,line width=0.25mm,white}, line width=0.5mm, black, fill=white, rectangle, minimum height=3em, minimum width=5em]
\tikzstyle{sum} = [draw,circle]
\tikzstyle{branch} = [circle,inner sep=0pt,minimum size=1mm,fill=black,draw=black]
\tikzstyle{nvbranch} = [circle,inner sep=0pt,minimum size=1mm,fill=white,draw=white, fill opacity=0, draw opacity=0]
\tikzstyle{vecBranch} = [circle,inner sep=0pt,minimum size=2mm,fill=black,draw=black]
\tikzstyle{input} = [coordinate]
\tikzstyle{output} = [coordinate] 
\tikzstyle{coord} = [coordinate] 
\tikzstyle{pinstyle} = [pin edge={to-,thin,black}] 
\tikzstyle{vecArrow} = [thick, decoration={markings,mark=at position
   1 with {\arrow[semithick]{open triangle 60}}},
   double distance=1.4pt, shorten >= 5.5pt,
   preaction = {decorate},
   postaction = {draw,line width=1.4pt, white,shorten >= 4.5pt}]
\tikzstyle{vecWithoutArrow} = [thick,
   double distance=1.4pt,
   postaction = {draw,line width=1.4pt, white}]
\tikzset{
  Pfeil/.style={thick,shorten >=#1,shorten <=#1,->,>=latex}, 
  UPfeil/.style={black,Pfeil=#1,font={\sffamily\itshape}},
  IPfeil/.style={black,Pfeil=#1,font={\ttfamily\itshape}} 
}

\newacronym{dde}{DDE}{delay-differential equation}
\newacronym{pde}{PDE}{partial-differential equation}
\newacronym{ode}{ODE}{ordinary-differential equation}
\newacronym{dpde}{DPDE}{delay-partial-differential equation}
\newacronym{dpde1}{D(P)DE1}{delay-partial-differential equation}
\newacronym{dpde5}{D(P)DE5}{delay-partial-differential equation}
\newacronym{fdm}{FDM}{finite difference method}
\newacronym{bvp}{BVP}{boundary value problem}
\newacronym{rms}{RMS}{root-mean-square}

\journal{Journal of Heat and Mass Transfer}

\setbool{markChanges}{false}

\begin{document}

\begin{frontmatter}

\title{On delay-partial-differential and delay-differential thermal models for variable pipe flow}

\author[1]{Jens Wurm}
\ead{jens.wurm@umit.at}

\author[1]{Simon Bachler}
\ead{simon.bachler@umit.at}

\author[1]{Frank Woittennek}
\ead{frank.woittennek@umit.at}

\address[1]{Institute of Automation and Control Engineering, \\ 
			University for Health Sciences, Medical Informatics and Technology, \\
		    Eduard Wallnöfer Zentrum 1, Hall in Tirol, Austria}

\begin{abstract}
  A new formulation of physical thermal models for variable plug flow through a pipe
  is proposed. The derived model is based on a commonly used one-dimensional distributed parameter
  model, which explicitly takes into account the heat capacity of the jacket of the pipe. The main result of the present contribution is the 
  constitution
  of the equivalence of this model with a serial connection of a pure delay or transport system and another \gls{pde}, subsequently called 
  \gls{dpde}-model. 
  The means for obtaining the proposed model comprise operational calculus in the Laplace domain as well as classical theory of characteristics.
  The finite-dimensional approximation of the \gls{dpde}-model leads to a 
  \gls{dde}-system, which can be seen as a generalization of commonly used 
  \gls{dde}-models consisting of a first-order low-pass filter subject to an input 
  delay.   
  The proposed model is compared to several alternative models in
  simulations and experimental studies. 
\end{abstract}


\end{frontmatter}

\begin{table*}[h]
\begin{framed}
    \nomenclature{$t$}{time \nomunit{\si{\second}}}
    \nomenclature{$z$}{spatial coordinate \nomunit{\si{\meter}}}
    \nomenclature{$r$}{radial coordinate \nomunit{\si{\meter}}}
    \nomenclature{$\upOmega$}{domain}

    \nomenclature[C]{$\QA$}{cross section surface \nomunit{\si{\square\meter}}}
    \nomenclature[C]{$U$}{perimeter \nomunit{\si{\meter}}}
    \nomenclature[C]{$\lambda$}{thermal conductivity \nomunit{\si{\watt\per\meter\per\kelvin}}}
    \nomenclature[C]{$\alpha$}{heat transfer coefficient \nomunit{\si{\watt\per\square\meter\per\kelvin}}}
    \nomenclature[C]{$c_{\mathrm{p}}$}{heat capacity 
    \nomunit{\si{\joule\per\kilo\gram\per\kelvin}}}
    \nomenclature[C]{$R$}{radius \nomunit{\si{\meter}}}
    \nomenclature[C]{$l$}{length of pipe \nomunit{\si{\meter}}}
    \nomenclature[C]{$A$}{surface area \nomunit{\si{\square\meter}}}
    \nomenclature[C]{$m$}{mass \nomunit{\si{\kilogram}}}
    \nomenclature[C]{$\epsilon$}{correction factor}
    \nomenclature[C]{$\rho$}{density \nomunit{\si{\kilogram\per\cubic\meter}}}

    \nomenclature[F]{$v$}{velocity \nomunit{\si{\meter\per\second}}}
    \nomenclature[F]{$T$}{temperature \nomunit{\si{\celsius}}}
    \nomenclature[F]{$\tau$}{transport delay time \nomunit{\si{\second}}}
    \nomenclature[F]{$\dq$}{heat flux \nomunit{\si{\watt\per\square\meter}}}

    \nomenclature[I]{m}{medium}
    \nomenclature[I]{w}{wall}
    \nomenclature[I]{$\infty$}{ambient}
    \nomenclature[I]{s}{shell}
    \nomenclature[I]{ma}{medium-ambient}
    \nomenclature[I]{mw}{medium-wall}
    \nomenclature[I]{wa}{wall-ambient}
    \nomenclature[I]{in}{input}
    \nomenclature[I]{out}{output}
    \nomenclature[I]{del}{delayed}
\printnomenclature
\end{framed}
\end{table*}
\glsreset{pde}
\glsreset{dde}
\glsreset{ode}
\glsreset{dpde}

\section{Introduction}

Plug flow models are widely used in various applications to describe the 
thermal behavior of
fluid flows through long pipes. Particular examples comprise solar desalination
plants \cite{Santos2011}, district heating grids \cite{Jie2012}, the thermal
behavior of catalysts \cite{Qiu2014, Lepreux2009, Lepreux2009model}, cooling loops of large 
gas engines \cite{Bachler2017b}, and solar thermal plants \cite{Cirre2007}. Commonly, one-dimensional \glspl{pde} or \glspl{dde}
are used for these purposes. While the \gls{pde}-models are based directly on
the mathematical description of the transport phenomena in combination with the
heat exchange between fluid and wall, the \gls{dde}-models 
are obtained heuristically by augmenting simple physically motivated 
\gls{ode}-models with additional delays in order to account for the transport 
phenomenon \cite{Kicsiny2014}. 
A physically based modeling approach for constant flow rate is presented in 
\cite{Lepreux2009, Lepreux2009model} describing the thermal behavior of an oxidation catalyst. Moreover, an approximation of the \gls{pde}-model 
by an corresponding 
diffusion equation is discussed in \cite{Lepreux2012}. In contrast a data driven model is introduced by \cite{Kicsiny2017}.
To the best knowledge of the authors no consistently physically based 
\gls{dde}-modeling approach has been published yet for the variable flow case. 
However, thanks to their simple structure \gls{dde}-models have been proven to be 
well suited in control applications \cite{Pekar2017, Bekiaris-Liberis2013, 
Mounier1998}.

These observations motivate the derivation of generalized
\gls{dde}-models, which constitute one of the main results of the present
work. The second contribution consists in an alternative \gls{pde}-model, which
separates the transport process from the filtering dynamics of the pipe.

Starting from a detailed one-dimensional \gls{pde}-model for the fluid inside 
the pipe and a two-dimensional \gls{pde}-model for the wall an 
one-dimensional \gls{pde}-model is derived. The latter approximates both 
the transport process of the fluid and the dynamics of the heat exchange 
between wall and fluid. It constitutes the basis of all models developed in the 
present 
contribution.
Originating from this model the well known \gls{ode}-model is derived, 
which can be intuitively adapted to the common used standard \gls{dde}-approach.
The new pipe model proposed in the present contribution is also based on the 
one-dimensional \gls{pde}-model. It combines a \gls{pde}- and \gls{dde}-model 
and is named \gls{dpde}-model due to its particular structure and is suitable for both
constant and variable flow rate. An approximation of the new \gls{dpde}-model reveals 
the new generalized \gls{dde}-models with a similar structure as the common 
used one. Finally, the different models are compared against each other and 
validated by measurements.

The contribution is structured as follows: The standard pipe models are 
introduced in \secref{sec:pipemodels}. The new pipe model is presented and 
linked to the standard approaches in \secref{sec:filter_model}. In 
\secref{sec:approxDDE} the general \gls{dde}-approaches are derived,
analyzed and optimized. All models are compared by simulation studies and validated 
by measurements in \secref{sec:validation}.

\section{Pipe Models} \label{sec:pipemodels}
\begin{figure}[tb]
	\centering
	\begin{tikzpicture}[auto, >=latex', scale=0.65, every node/.style={scale=0.85}]

    \filldraw[fill=gray!40, draw=black] (-3, 7) rectangle (-2.25, 8);
    \filldraw[fill=gray!40, draw=black] (-0.25, 7) rectangle (16, 8);

    \filldraw[fill=gray!40, draw=black] (-2.25, 6.75) -- (-0.25, 8.25) -- (-0.25, 6.75) -- (-2.25, 8.25) -- (-2.25, 6.75);

    \filldraw[fill=gray!40, draw=black] (0.75, 6.5) rectangle (1.25, 8.75);
    \filldraw[fill=gray!40, draw=black] (0.5, 6.5) rectangle (1.5, 8.5);

    \filldraw[fill=gray!40, draw=black] (4.6+0.75, 6.5) rectangle (4.6+1.25, 8.75);
    \filldraw[fill=gray!40, draw=black] (4.6+0.5, 6.5) rectangle (4.6+1.5, 8.5);

    \filldraw[fill=gray!40, draw=black] (9.2+0.75, 6.5) rectangle (9.2+1.25, 8.75);
    \filldraw[fill=gray!40, draw=black] (9.2+0.5, 6.5) rectangle (9.2+1.5, 8.5);

    \filldraw[fill=gray!40, draw=black] (14.75, 6.5) rectangle (15.25, 8.75);
    \filldraw[fill=gray!40, draw=black] (14.5, 6.5) rectangle (15.5, 8.5);

	\filldraw[left color = olive!50,right color = olive!10] (0,0.5) rectangle (8,3.5);
	\filldraw[fill=gray!40, draw=black] (0, 0) rectangle (8, 0.5);
	\filldraw[fill=gray!40, draw=black] (0, 3.5) rectangle (8, 4);

    \draw[black, dashed] (1.8, 6.75) rectangle (4.8, 8.25);

    \draw[black, dashed] (1.8, 6.75) -- (0, 4);
    \draw[black, dashed] (4.8, 6.75) -- (8, 4);

    \draw[-] (0, -0.2) -- (0, -0.4);
    \draw[->] (0, -0.3) -- node[below] {$z$} (1, -0.3);

    \draw[->] (4, 1.25) -- (5.5, 1.25);
    \draw[->] (4, 1.75) -- (5.5, 1.75);
    \draw[->] (4, 2.25) -- (5.5, 2.25);
    \draw[->] (4, 2.75) -- (5.5, 2.75);

    \draw[->] (-1.75, 1.25) -- (-0.25, 1.25);
    \draw[->] (-1.25, 1.75) -- (-0.25, 1.75);
    \draw[->] (-1.25, 2.25) -- (-0.25, 2.25);
    \draw[->] (-1.75, 2.75) -- (-0.25, 2.75);

    \draw (-2, 2) node {$\Tmin(t)$};

    \draw[->] (1.5, 3) -- (1.5, 3.5);
    \draw[->] (2, 3) -- (2, 3.5);
    \draw[->] (2.5, 3) -- (2.5, 3.5);
    \draw[->] (3, 3) -- (3, 3.5);
    \draw (2.25, 2.75) node {$\dq(z,t)$};
    \draw (0.5, 3.25) node {$\cdot$};
    \draw (1, 3.25) node {$\cdot$};
    \draw (3.5, 3.25) node {$\cdot$};
    \draw (4, 3.25) node {$\cdot$};

    \draw[-] (1, 9) -- (1, 9.2);
    \draw[-] (15, 9) -- (15, 9.2);
    \draw[<->] (1, 9.1) -- node [above] {$l$} (15, 9.1);

    \draw[fill=gray!40, draw=black] (4+10, 2) circle [radius=2];
    \draw[fill=olive!50, draw=black] (4+10, 2) circle [radius=2-0.5];

    \draw[-]  (7.5, 0.25+1) -- (7.75, 0.1+1) -- node[above] {$\Tm(z,t)$} (13.75, 0.1+1) -- (14, 0.25+1);
    \draw[-]  (7.5, 0.25) -- (7.75, 0.1) -- node[above] {$\Tw(z,t)$} node[below] {$\Ta(t)$} (13.75, 0.1) -- (14, 0.25);

    \draw[<->] (4+10, 2) -- node [above] {$\Rm$} (5.5+10, 2);
    \draw[<->] (4+10, 2) -- node [left] {$\Rw$} (4+10, 4);
\end{tikzpicture}
	\caption{Sketch of pipe test rig and used variables.}
	\label{fig:general_pipe}
\end{figure}
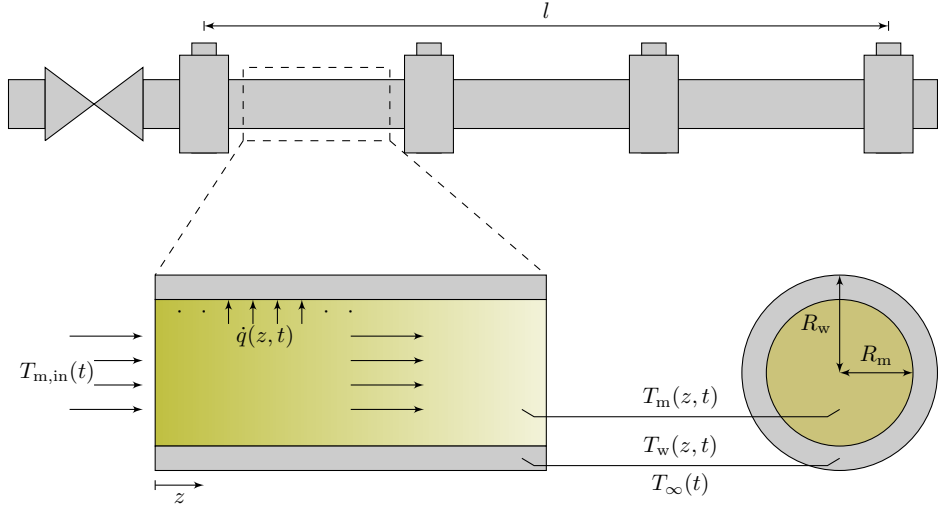

Two models of different type describing the thermal behavior of a plug 
flow in a pipe are presented within this section.
\figref{fig:general_pipe} shows the considered pipe of length $l$ with 
inner and outer radii  $\Rm$ and $\Rw$. The medium temperature 
is denoted by $\Tm$, $\Tw$ is the wall temperature, and $\Ta$ describes the 
ambient temperature. The input temperature profile is $\Tmin$, the velocity of 
the medium is $\vm$, and $\dq$ stands for the heat flux between medium, wall, 
and ambient.
For the modeling an 
\begin{enumerate*}[label=(\roman*)]
	\item incompressible medium with\label{ass:incrompres}
	\item a radially constant temperature and velocity profile due to turbulent flow is 
	assumed. Furthermore, \label{ass:turbulent} thanks to a sufficiently large medium 
	velocity  
	\item the thermal conduction in flow direction is neglected for both the 
        wall and the medium\label{ass:neg_themal_conduction}, and
	\item all material parameter are assumed to be spatially and temporally 
	constant.\label{ass:param_const}
\end{enumerate*}

\subsection{Partial-differential equation-approach}
For the sake of simplicity\footnote{Note that the further findings and calculations 
are not restricted to the cylindrical geometry but can be adapted to arbitrary cross-sections. However, the generalization would require some
notational and computational effort.} the pipe is assumed cylindrical with cross section $\QAp\subset\Re^2$. 
In the following the equations are written in 
cylindrical coordinates with  $r$ the radial coordinate, $z$ the axial coordinate. The angular coordinate is dropped due to a symmetry assumption. 
The pipe shown in \figref{fig:general_pipe} can be separated into a medium 
and a wall part, which can be treated separately.

\subsubsection{Medium}
Taking  into account Assumption \ref{ass:turbulent}, a constant 
temperature profile over the cross section $\QAm$ of the medium can be assumed. 
Moreover, considering Assumption \ref{ass:neg_themal_conduction} leads to the 
well known one-dimensional transport-equation describing the fluid flow in 
$z$-direction (see, e.g., \cite{Witelski2015}):
\begin{align}
    \Am\cpm\rhom\left(\pderit\Tm(z,t) + \vm(t)\pderiz\Tm(z,t)\right) & = 
    -2\pi \Rm\dq(\Rm,z,t),	\label{eqn:pde_medium_1d}
\end{align}
with the specific heat capacity $\cpm$ and the density $\rhom$ of the medium.  
The heat flux $\dq(\Rm,z,t)$ from the medium into the wall is detailed below in \eqref{subeqn:bc_mw}.
Moreover, the corresponding inflow boundary condition and the initial condition are given by
\begin{alignat*}{2}
    \Tm(0,t) &= \Tmin(t),& \qquad
	\Tm(z,0) & = \Tm[0](z),
\end{alignat*}
respectively, with the input temperature $\Tmin$ and the initial temperature profile $z\mapsto\Tm[0](z)$.

\subsubsection{Wall}
The evolution of the temperature distribution within the wall is described by 
the heat equation \cite[p. 87]{Bergman2011}, which reads in cylindrical 
coordinates:
\begin{align}
    \rhow\cpw\pderit\Tw(r, z,t) & =  -\frac{1}{r}\pderir\left(r 
    \dq(r,z,t)\right),\, \dq(r,z,t)  =-\lambdaw\pderir 
    \Tw(r,z,t). \label{eqn:pde_wall_3d}
\end{align}
Here $\lambdaw$ denotes the thermal conductivity of the wall and $\dq$ is the 
radial component of the heat flux within the wall. 
Above, the possible dependency of the wall temperature on an angular coordinate has been dropped for symmetry reasons. Moreover, the heat flux in axial direction has been neglected in view of
Assumption \ref{ass:neg_themal_conduction}.

The boundary conditions for the shell surface are given by Fourier's Law
\begin{subequations}\label{eqn:pde_wall_q}
\begin{alignat}{3}
    -\dq(\Rw,z,t)& = \lambdaw\pderir\Tw(\Rw,z,t)&&=& -{\alphawa}&(\Tw(\Rw,z,t) -  \Ta(t)), \label{subeqn:bc_wa}\\
    -\dq(\Rm,z,t) & =\lambdaw\pderir\Tw(\Rm, z,t)&&=& {\alphamw}&(\Tw(\Rm,z,t) - \Tm(z,t). \label{subeqn:bc_mw}
\end{alignat}
\end{subequations}
In the above equations 
linear heat transfer between the wall and the medium respectively the medium and the ambient is assumed. The 
respective heat transfer coefficients are denoted by
$\alphamw$ and $\alphawa$.
Finally, the initial conditions read
\begin{align*}
	\Tw(r,z,0) & = \Tw[0](r,z).
\end{align*}

\subsubsection{Overall one-dimensional model}\label{subsubsec:overall_1d_model}
The complete one-dimensional model is derived by combining the models for the medium and the wall. To this end, the wall temperature model is reduced
to a one-dimensional model by averaging the wall temperature over the area $\Aw=(\Rw^2 - \Rm^2)\pi$ of the cross sectional surface $\QAw$:
\begin{align}
    \barTw(z,t) & = \frac{2\pi}{\Aw}\int\limits_{\Rm}^{\Rw}\Tw(r,z,t) r\dd 
    r. \label{eqn:pde_wall_1d_mean}
\end{align}
Similarly, integrating the \gls{pde}-model of the wall temperature 
\eqref{eqn:pde_wall_3d} over $\QAw$ yields  
\begin{align*}
	\Aw\rhow\cpw\pderit\barTw(z,t) & = 2\pi\lambdaw\Bigl[
	r\pderir\Tw(r,z,t)\Bigr]_{\Rm}^{\Rw}\\
    & = 2\pi\lambdaw\Bigl[ \Rw\pderir\Tw(\Rw,z,t)-\Rm\pderir\Tw(\Rm,z,t)\Bigr]
\end{align*}
with the averaged wall temperature given by \eqref{eqn:pde_wall_1d_mean}.
Substituting the boundary derivatives on the right hand side by the boundary 
conditions 
\eqref{eqn:pde_wall_q} reveals
\begin{align*}
\begin{split}
	\Aw\rhow\cpw\pderit\barTw(z,t)
 = 2\pi\Bigl[\Rw&\alphawa\bigl(\Ta(t) - \Tw(\Rw, z, 
 t)\bigr)\\+\Rm&\alphamw\bigl(\Tm(z,t) - \Tw(\Rm,z,t)\bigr)\Bigr]. 
\end{split}
\end{align*}
Finally, the boundary wall temperature is approximated by the average 
temperature \eqref{eqn:pde_wall_1d_mean}.
This leads to the one-dimensional model
\begin{multline*} 
	\Aw\rhow\cpw\pderit\barTw(z,t)\\
 = \Uw\balphawa\bigl(\Ta(t) - \barTw(z, t)\bigr)+\Um\balphamw\bigl(\Tm(z,t) - 
 \barTw(z,t)\bigr)\end{multline*}
for the pipe jacket, where the new overall heat transfer coefficients 
$\balphamw$ and $\balphawa$ are defined by
\begin{align*}
    \frac{1}{\balphamw} & = \frac{1}{\alphamw} + \frac{\barRm}{\lambdaw}, \\
	\frac{1}{\balphawa} & = \frac{1}{\alphawa} + \frac{\barRw}{\lambdaw},
\end{align*}
with
\begin{align*}
	\barRm &= \Rm\left(\frac{\Rw^2}{\Rw^2-\Rm^2}\ln\left(\frac{\Rw}{\Rm}\right) 
	- \frac12\right),\\
	\barRw &= \Rm \left(-\frac{\Rm^2}{\Rw^2 - \Rm^2} 
	\ln\left(\frac{\Rw}{\Rm}\right) + \frac12\right)
\end{align*}
for a cylindrical pipe profile. They are chosen in such a way, that the substitution 
is exact in the stationary regime (cf.\ \ref{app:overallcoefficients}).
Moreover, the perimeters $\Um = 2 \pi \Rm$ and $\Uw = 2 \pi \Rw$ are introduced 
for ease of notation.
The same approximation is applied to the heat flux \eqref{subeqn:bc_mw} appearing on the right hand side of the one-dimensional \gls{pde}-model \eqref{eqn:pde_medium_1d} for the medium temperature. Thus, \eqref{eqn:pde_medium_1d} can be rewritten as
\begin{align*}
	\cpm\rhom&\left(\pderit\Tm(z,t) + \vm(t)\pderiz\Tm(z,t)\right) = 
    \frac{\Um}{\Am}\balphamw\left(\barTw(z,t) - \Tm(z,t)\right).
\end{align*}

Finally, the thermal behavior of a plug flow through a pipe can be described by 
the one-dimensional \gls{pde}-system
\begin{subequations}
    \label{eqn:pde_1d}
\begin{align}
\vm(t) \pderiz \Tm(z,t) + \pderit \Tm(z,t) & = h_1 (\Tw(z,t) - \Tm(z,t)) 
\label{subeqn:pde_1d_Tm}\\
\begin{split}
\pderit \Tw(z,t) & = h_2 \left(\Tm(z,t) - \Tw(z,t)\right) \\
                 & \quad -  h_3\left(\Tw(z,t) - \Ta(t)\right)
\end{split} \label{subeqn:pde_1d_Tw}
\end{align}
with boundary condition
\begin{align}
\Tm(0,t) & = \Tmin(t) \label{eqn:pde_1d_boundary}
\end{align}
and the initial conditions
\begin{alignat}{2}
\Tm(z,0) & = \Tm[0](z),\quad & \Tw(z,0) & = \Tw[0](z) \label{eqn:pde_1d_ic}.
\end{alignat}
\end{subequations}
The physical parameters are collected in
\begin{align*}
	h_1 & = \frac{\Um}{\Am}\frac{\balphamw}{\rhom\cpm},& h_2 & = 
	\frac{\Um}{\Aw}\frac{\balphamw}{\rhow\cpw}, & h_3 & = 
	\frac{\Uw}{\Aw}\frac{\balphawa}{\rhow\cpw}.
\end{align*}
Therein and below the averaged wall temperature $\bar{T}_w$ is denoted by $\Tw$  for notational simplicity.

\begin{remark*}
Since the convection boundary layer between the medium and the 
wall varies at different velocities the heat transfer 
coefficients may dependent on velocity \cite{Bergman2011}. Hence, the \gls{pde}-model 
\eqref{eqn:pde_1d} can be extended to
\begin{align*}
    \vm(t) \pderiz \Tm(z,t) + \pderit \Tm(z,t) & = h_1\left(\vm(t)\right) \left(\Tw(z,t) - \Tm(z,t)\right)\\
    \begin{split}
        \pderit \Tw(z,t) & = h_2\left(\vm(t)\right) \left(\Tm(z,t) - \Tw(z,t)\right) \\
                         & \quad -  h_3\left(\Tw(z,t) - \Ta(t)\right).
    \end{split}
\end{align*}
Assuming an affine velocity dependence of $\balphamw$ with the slope $\alphamw[1]$ (unit
$\si[per-mode=symbol]{\watt\second\per\cubic\meter\per\kelvin}$) and the intercept $\alphamw[0]$ (unit
$\si[per-mode=symbol]{\watt\per\square\meter\per\kelvin}$) the heat transfer coefficients are given by:
\begin{align*}
    h_1\left(\vm(t)\right) & = \frac{\Um}{\Am}\frac{\alphamw[0] + 
    \alphamw[1]\vm(t)}{\rhom\cpm},\\ 
    h_2\left(\vm(t)\right) & = \frac{\Um}{\Aw}\frac{\alphamw[0] + 
    \alphamw[1]\vm(t)}{\rhow\cpw}.
\end{align*}
\end{remark*}

\subsection{Ordinary-differential equation-approach}
If the output temperature of the pipe is of particular interest and transport
delays do not play a significant role simple
\gls{ode}-models can be employed instead of the above derived \gls{pde}. Such models are preferred for example for automotive cooling loops
(cf.\ \cite{Aschemann2011}). The derivation of the model equations starting from
\eqref{eqn:pde_medium_1d} is sketched below.

In contrast to the presented  
\gls{pde}-models, the dynamics of the wall temperature is not  explicitly taken 
into account. Hence, a heat flux 
\begin{align}
	\dot{q}(z,t) & = -\alphama(\Tm(z,t) - \Ta(t)) \label{eqn:alphama}
\end{align}
is observed between medium and ambient instead of \eqref{eqn:pde_wall_q}, 
with an overall heat transfer coefficient (cf.\ \cite[p.~31~ff.]{Baehr2011})
\begin{align}
	\frac{1}{\alphama} &= \frac{1}{\alphamw} + \frac{1}{\alphawa} + 
	\frac{\Rm}{\lambdaw} \ln\left(\frac{\Rw}{\Rm}\right). \label{eqn:alpha_ma}
\end{align}
Thus, the simplified one-dimensional \gls{pde}-model gets\footnote{Note that the heat capacity of the wall can be accounted for by an additional
coefficient in front of $\pderit\Tm(z,t)$ (see \eqref{eqn:adapted_common_dde})}
\begin{align}
    \cpm\rhom\left[\vm(t)\pderiz\Tm(z,t) + \pderit \Tm(z,t)\right] & = 
    \frac{\Uw}{\Am}\alphama (\Ta(t) - \Tm(z,t)), \label{eqn:pde_wo_wall}
\end{align}
with the boundary condition
\begin{align*}
	\Tm(0,t) = \Tmin(t)
\end{align*}
and the initial condition
\begin{align*}
    \Tm(z,0) = \Tm[0](z).
\end{align*}
The simplified \gls{pde}-model \eqref{eqn:pde_wo_wall} can be interpreted as a 
further approximation of the \gls{pde}-model \eqref{eqn:pde_1d}, where the 
dynamics of the wall temperature are neglected and the new overall heat 
transfer coefficient \eqref{eqn:alpha_ma} is derived based on the stationary 
wall temperature profile.

In a further approximation step a spatial discretization of \eqref{eqn:pde_wo_wall} with the simple difference quotient
\begin{align*}
    \left(\pderiz \Tm\right)\left(l, t\right) & \approx \frac{1}{l}\left(\Tm(l, t) - \Tmin(t)\right)
\end{align*}
leads to the \gls{ode} (with $\Tm(t):=\Tm(l,t)$)
\begin{align}
    \derit \Tm(t) &= \frac{\vm(t)}{l} \left( \Tmin(t) - \Tm(t) \right) + h_4 \left (\Ta(t) - \Tm(t) \right)  \label{eqn:ode:approach}
\end{align}
describing the average medium temperature of the pipe. Therein
\begin{align*}
    h_4 & = \frac{\Uw}{\Am} \frac{\alphama}{\cpm\rhom}.
\end{align*}
Due to constant material parameters (cf.\ Assumption \ref{ass:param_const})
\eqref{eqn:ode:approach} can easily be generalized to
\begin{align*}
    \derit \Tm(t) &= \frac{\vm(t)}{l} \left( \Tmin(t) - \Tm(t) \right) + \frac{\alphama\, \As}{\cpm \mmed} \left (\Ta(t) - \Tm(t) 
    \right),
\end{align*}
where $\mmed$ describes the mass of the medium inside the pipe and $\As$ the shell surface area. \footnote{Note, that in 
a system theoretical sense the \gls{ode} representation of the pipe equals a filtering of the input and ambient temperature with a first-order 
low-pass filter.}
The latter model is used for configurations allowing for the neglection of the transport phenomenon, e.g. in automotive cooling loops, where the 
pipes are rather short 
\cite{Aschemann2011}.

\subsection{Delay-differential equation-approach} 
\label{subsec:dde_approach}
If the transport delays within the medium cannot be neglected, as in solar field 
applications \cite{Cirre2007, Santos2011} or systems with long 
pipes \cite{Pekar2017, Bachler2017b}, 
the simple \gls{ode}-approach \eqref{eqn:ode:approach} is intuitively 
complemented by the variable transport delay $\tau$, implicitly defined by:
\begin{align*}
    \int\limits_{t-\tau(t)}^{t} \vm(\zeta) \, \dd \zeta & = l.
\end{align*}
This way, one obtains the \gls{dde}-model
\begin{align}
    \pderit \Tm(t) &= \frac{\vm(t)}{l} \left( \Tmin(t - \tau(t)) - \Tm(t) \right) + h_4 \left (\Ta(t) - \Tm(t)  \right) \label{eqn:dde:approach}
\end{align}
as discussed in \cite{Kicsiny2014}.

\begin{figure}[tb]
	\centering
	\begin{tikzpicture}[auto, >=latex', scale=0.65, every node/.style={scale=0.85}]

    \fill[fill=gray!40] (0, 0) rectangle (14, 1);
    \fill[fill=gray!40] (10, -2) rectangle (13, 3);

    \draw[black, thick] (0, 1) -- (10, 1) -- (10, 3) -- (13, 3) -- (13, 1) -- (14, 1);
    \draw[black, thick] (0, 0) -- (10, 0) -- (10, -2) -- (13, -2) -- (13, 0) -- (14, 0);

    \filldraw[fill=black] (11.4, -0.5) rectangle (11.6, 3.5);
    \filldraw[fill=black] (11.1, -0.5) ellipse (0.5 and 0.2);
    \filldraw[fill=black] (11.9, -0.5) ellipse (0.5 and 0.2);

    \draw[black, ->, thick] (11.325, 3.25) arc [start angle=255, end angle=-75, x radius=0.75, y radius=0.25];

    \draw (5.5, 2) node {$\Ta(t)$};
    \draw (11.5, -2) node [above] {$\Tm(t)$};
    \draw (0, 0.5) node [right] {$\Tmin(t)$};
    \draw (10, 0.5) node [left] {$\Tmin(t-\tau)$};

    \draw[black, ->] (-0.5, 0.5) -- (0, 0.5);
    \draw[black, ->] (14, 0.5) -- (14.5, 0.5);

\end{tikzpicture}
	\caption{Physical interpretation of the classical \gls{dde}-pipe model.}
	\label{fig:dde:approach}
\end{figure}
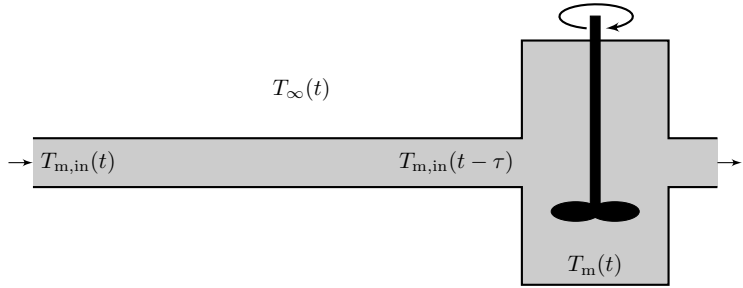

Though such models have been proven to be useful in applications within their derivation the transport phenomenon is considered 
twice: After approximating the
transport equation by means of a first-order \gls{ode} and  abandon the transport delay, the latter will be introduced again in a consecutive modeling step.
A physical interpretation of the obtained \gls{dde}-model for a constant velocity is 
depicted in \figref{fig:dde:approach}.
It shows an ideal pipe (pure convection) connected to an ideally stirred tank, which 
models the heat dissipation as well as the heat capacity of the wall.
However, at first glance the separation of the transport process and the dynamics does not seem reasonable.
In \cite{Bachler2017a} it is shown that a \gls{pde}-approach, which does not 
explicitly consider the heat capacity of the wall  does not reveal a pipe model like 
\eqref{eqn:dde:approach}.
Nevertheless, the heat capacity of the wall can be considered heuristically by extending the \gls{dde}-model \eqref{eqn:dde:approach} 
with an additional correction factor $\epsilon$ to
\begin{align}
	\epsilon\,\pderit \Tm(t) &= \frac{\vm(t)}{l} \left(\Tmin(t-\tau(t)) - 
	\Tm(t) \right) + \frac{\alphama \As}{\cpm\mmed} \left (\Ta(t) - \Tm(t) \right), 
	\label{eqn:adapted_common_dde}
\end{align}
as proposed in \cite{Bachler2017a}.
However, when explicitly taking into account the heat capacity of the wall 
similar results may be obtained by a first order approximation of the transfer function in the frequency domain for 
constant flow rates \cite{Lepreux2009, Lepreux2009model}.

\section{Delay-Partial-Differential-Equation Model} \label{sec:filter_model}
The new modeling approach described below is based on the one-dimensional \gls{pde} 
\eqref{eqn:pde_1d} for plug flow through a pipe with additional heat storage within  
the wall and heat transfer between  
medium and wall respectively wall and ambient. Due to its particular structure 
separating the transport phenomena within the medium from the wall dynamics the 
new model is named \gls{dpde}-model in the following.
In a first step the basic ideas are sketched under the simplifying assumptions 
of a constant flow rate and a perfectly isolated pipe in the Laplace domain.
Therefore, the transfer function of the one-dimensional pipe model  
\eqref{eqn:pde_1d} is analyzed. This part basically restates the results already presented in \cite{Lepreux2009, Lepreux2009model}. 
Based on that findings the new \gls{dpde}-model is introduced in form of a second order \gls{pde} with delayed boundary condition.
Afterwards, this formulation is generalized by taking into account the heat loss to the 
ambient and time dependent flow velocities.
\subsection{Constant flow rate and perfect isolation}
Assuming a constant flow rate $\vm$ of the fluid, perfect isolation ($h_3=0$), and homogeneous initial
conditions\markChange{\footnote{\markChange{At this point we are primarily interested in the input-output
behaviour, i.e., in computing the transfer function. Consequently inhomogeneous initial conditions can be assumed without loss of generality. In
contrast the \gls{dpde}-model computed at the end of the subsection constitutes a particular realization of this transfer function only. However, the time
domain computations in \secref{sec:variable_flow_rate} show that this \gls{dpde}-model is indeed equivalent to the original
\gls{pde} description.}}},
$\Tm(z,0)=\Tw(z,0)=0$, the \gls{bvp} \eqref{eqn:pde_1d} can be transformed into the Laplace domain
\begin{subequations} \label{eqn:pde_laplace}
\begin{align}
    \vm \pderiz \hatTm(z,s) + s \hatTm(z,s) & = h_1 \left(\hatTw(z,s) - \hatTm(z,s)\right) \label{subeqn:pde_laplace_medium}\\
\begin{split}
    s\hatTw(z,s) & = h_2 \left(\hatTm(z,s) - \hatTw(z,s)\right) \label{subeqn:pde_laplace_wall}
\end{split}
\end{align}
\end{subequations}
with the Laplace transforms $\hatTm$ and $\hatTw$ of 
the corresponding temperatures \cite{Lepreux2009, Lepreux2009model}.
Eliminating the wall temperature from \eqref{eqn:pde_laplace}  
 one obtains
\begin{equation}
    \vm\pderiz\hatTm(z,s) = \zeta(s)\hatTm(z,s),\quad     \zeta(s)  = \frac{h_1 
    h_2}{s + h_2} -h_1 - s. \label{eqn:pde_subst_laplace:simplified}
\end{equation}
The solution of \eqref{eqn:pde_subst_laplace:simplified} is given by
\begin{equation*}
    \hatTm(z,s) = G_{\Tm}(z,s)\hatTm(0,s)\label{eqn:filter_Tm_Ta_sol_lap},\quad 
    G_{\Tm}(z,s) = \exp\left(\frac{z}{\vm}\zeta(s) \right).
\end{equation*}
The transfer function $G_{\Tm}(z,s)$ can be easily split up into three 
multiplicative parts given by
  \begin{equation*}
    G_1(z,s) \!=\! \exp\left(-s\frac{z}{\vm} \right),\;
    G_2(z)  \!= \!\exp\left( - h_1\frac{z}{\vm} \right),\;
    G_3(z,s)  \!=\! \exp\left(\frac{z}{\vm}\frac{h_1h_2}{s + h_2} \right).
  \end{equation*}
  Therein, $G_1(s)$ corresponds to a pure spatially dependent time delay and
  $G_2(s)$ is a spatially dependent scaling factor. Moreover, as explained below, $G_3(z,s)$ can be interpreted as an 
infinite-dimensional filter without any time delay. As a consequence, the overall structure of the transfer function is similar to 
the structure of the \gls{dde}-model \eqref{eqn:dde:approach} 
discussed in \secref{subsec:dde_approach}.
The main difference is constituted by the filter part $G_3(z,s)$ replacing the 
simple first-order low-pass filter in the \gls{dde}-model 
\eqref{eqn:dde:approach}. In order to 
compute the impulse response of this filter, i.e., transforming the 
input-output relation associated with $G_3(z,s)$ into the time domain, the 
transfer function is expanded into a power series:
\begin{align}
	G_3(z,s) & = 1 + \sum\limits_{n=1}^{\infty}\frac{1}{n!}\left(\frac{z}{\vm}\frac{h_1 h_2}{s + h_2}\right)^n \label{eqn:filter_eq_exp}.
\end{align}
The sum within the latter expression can be interpreted as a parallel 
connection of an infinite number of low-pass filters of increasing order.
Element-wise computation of the inverse Laplace transform of 
\eqref{eqn:filter_eq_exp} yields the desired, spatially dependent impulse 
response:
\begin{align}
    g_3(z,t) & = \delta(t) + \exp\left(-h_2t\right) 
    \sum\limits_{n=0}^{\infty}\frac{\left(Kz\right)^{n+1}t^n}{(n+1)!n!} \notag 
    \\
             & = \delta(t) + \exp\left(-h_2t\right)\sqrt{K\frac{z}{t}}\sum\limits_{n=0}^{\infty}\frac{1}{(n+1)!n!}\sqrt{K z t}^{2n+1}, \label{eqn:filter_timedomain}
\end{align}
with $K = \frac{h_1 h_2}{\vm}$ and the Dirac delta distribution $\delta$.
With the substitution $x = 2\sqrt{K z t}$ the infinite sum within the above 
expression corresponds to the well known series expansion of the modified 
Bessel function of first order: 
\begin{align*}
    x\mapsto \besselI_1(x) & = \sum\limits_{n=0}^{\infty}\frac{1}{(n+1)!n!}\left(\frac{x}{2}\right)^{2n+1}.
\end{align*}
As a result the impulse response \eqref{eqn:filter_timedomain} can be rewritten as (cf.\ \cite{Lepreux2009model})
\begin{align}
    g_3(z,t) & = \delta(t) + \exp\left(-h_2 t\right)\sqrt{\frac{z}{\vm}\frac{h_1h_2}{t}}\besselI_1\left(2\sqrt{\frac{h_1 h_2 z \,
    t}{\vm}}\right). \label{eqn:filter_IR_time}
\end{align}
With the above computed impulse response \eqref{eqn:filter_IR_time} in the time domain the input-output relation corresponding to
$G_{\text{F}}(z,s)=G_2(z)G_3(z,s)$ is given by the convolution ($\star$) of $g_{\text{F}}(z,t)=G_2(z)g_3(z,t)$ and the delayed input
$\TminDel(\bullet)$ by:
\begin{align}\label{eq:filter:TmDel_Laplace}
    \TmDel(z,\bullet) & =  g_{\text{F}}(z,\bullet) \star \TminDel(\bullet)
\end{align}
at a specific time $\bullet$. Therein, $(z,t)\mapsto\TmDel(z,t)$ and $t\mapsto\TminDel(t)$ can be seen as delayed temperature 
profiles, which coincide with $\Tm$ 
\begin{align}\label{eq:filter:TmDel}
    \TmDel(z, t) & = \Tm\left(z,t-\tau(l-z)\right),
\end{align}
respectively $\Tmin$ 
\begin{align} \label{eqn:filter:TmDelAtl}
    \TminDel(t) & = \TmDel(0,t) = \Tmin(t-\tau(l))
\end{align}
up to the spatially dependent transport delay imposed by $G_1(z,s)$, i.e.,
\begin{align}
    \tau(z) & =\frac{z}{\vm}.
\end{align}
Observe that at the outflow boundary $z=l$ the delayed temperature corresponds to the actual temperature:
\begin{align} \label{eqn:filter:outputCond}
    \Tm(l, t) & = \TmDel\left(l,t\right).
\end{align}
In view of a intended numerical implementation of the \gls{dpde}-model a 
realization of the transfer function \eqref{eq:filter:TmDel_Laplace} as a 
\gls{bvp} has to be derived. This can be either achieved by means of the 
substitution
\begin{align*}
    \hatTm(z, s) & = \exp(s\tau(l-z))\hatTmDel\left(z,s\right),\quad \hatTmin(s)=\exp(s\tau(l))\hatTminDel(s)
\end{align*}
in \eqref{eqn:pde_subst_laplace:simplified} 
or, equivalently, by differentiating the relation
\begin{align*}
    \hatTmDel(z,s) & =  G_{\text{F}}(z,s)\hatTminDel(s),& G_{\textrm{F}}(z,s) & = G_2(z)G_3(z,s)
\end{align*}
in the Laplace domain. Both approaches yield the ordinary \gls{bvp}
\begin{align}
    \vm(s + h_2)\pderiz\hatTmDel(z,s)+h_1s\hatTmDel(z,s) & = 0,& \hatTmDel(0,s) & = \hatTminDel(s).  \label{eqn:pde_subst_laplace:simplified:shifted}
\end{align}
Translating this relation into the time domain leads to the desired \gls{pde}
\begin{equation}
    \vm(\pderit + h_2)\pderiz\TmDel(z,t)+h_1\pderit\TmDel(z,t)=0  \label{eqn:pde_subst:simplified:shifted},
 \end{equation}
which, together with the delayed inflow boundary condition \eqref{eqn:filter:TmDelAtl} and the output equation \eqref{eqn:filter:outputCond}
constitutes the complete new \gls{dpde}-model under the given simplifying 
assumptions.
Note that, the presented ideas immediately generalize to the non isolated case.

\subsection{Variable flow rate}\label{sec:variable_flow_rate}
In case of variable flow rates the formal computations in the Laplace domain 
are not applicable due to the time variance of the system considered.
Nevertheless, as shown below the ideas generalize even to a time varying 
setting in a similar way. To this end 
the spatial dependence delay introduced within the previous section is replaced by time 
and spatial depending transport delay, which can be described by 
the integral equation (similar definition for pure transport processes can be 
found e.g.\ in \cite{Kicsiny2014, breschpietri2016}):
\begin{align}
    \int\limits_{t-\tau(l-z,t)}^{ t} \vm(\zeta) \dd\zeta & = 
	\int\limits_{z}^{l}\dd \nu = l - z. \label{eqn:variable_delay_time}
\end{align}
Therein, $\tau(l - z,t)$ denotes the time which a portion of fluid arriving at a certain time $t$ at the outflow $z=l$ has 
traveled from the
point $l - z$.
Similarly as in \eqref{eq:filter:TmDel} the delayed temperature
\begin{align*}
    \TDel(z, t) &= \T\left(z, t - \tau(l - z, t)\right) = \T\left(z, \varphi(z, t)\right)
\end{align*}
is introduced, where the abbreviation $\varphi(z,t)=t-\tau(l-z,t)$ has been used for convenience and $\ast$ may be replaced by 
$\text{m}$, $\text{w}$ or $\infty$. The temporal and spatial derivatives of these delayed quantities are given by
\begin{subequations}\label{eqn:del_part_der}
\begin{align}
    \pderit \TDel(z, t) & =  \pderit \varphi(z, t)  \pderiphi \T(z, \varphi(z, t)), \\
    \pderiz\TDel(z, t) & = \pderiz \T(z, \varphi(z, t)) + 
    \pderiz \varphi(z, t) \pderiphi \T(z, \varphi(z, t)).
\end{align}
\end{subequations}
Therein, the derivatives of $\varphi$ w.r.t.\ $z$ and $t$ follow by differentiating  \eqref{eqn:variable_delay_time} and the usage of the
Leibniz integral rule, i.e., from
\begin{subequations}\label{eqn:variable_delay_time_der}
\begin{align}
    \pderiz \int_{\varphi(z,t)}^{t} \vm(\zeta) \dd\zeta & = -\pderiz \varphi(z,t)\vmDel(z,t)  = -1,\\
    \pderit \int_{\varphi(z,t)}^{t} \vm(\zeta) \dd\zeta & = \vm(t) - \pderit \varphi(z, t) \vmDel(z,t) = 0
\end{align}
\end{subequations}
with the delayed velocity
\begin{align*}
\vmDel(z, t) & = \vm\left(\varphi(z, t)\right).
\end{align*}
Taking into account \eqref{eqn:variable_delay_time_der},  \eqref{eqn:del_part_der} 
can be simplified to
\begin{align*}
	\pderit \TDel(z, t) & =  \frac{\vm(t)}{\vmDel(z, t)} \pderiphi \T(z, 
	\varphi(z, t)), \\
	\pderiz\TDel(z, t) & = \pderiz \T(z, \varphi(z, t)) + 
	\frac{1}{\vmDel(z, t)} \pderiphi \T(z, \varphi(z, t)).
\end{align*}
Substitution of 
\eqref{eqn:del_part_der} into the delayed version of \eqref{eqn:pde_1d} yields
\begin{subequations}\label{eqn:del_T_part_der}
\begin{align}
    \vmDel(z,t) \pderiz \TmDel(z, t) & = h_1\left( 
    \TwDel(z,t) - \TmDel(z,t)\right) \label{eqn:del_Tm_part}\\
    \pderit \TwDel(z,t) & = \frac{\vm(t)}{\vmDel(z, t)}\Big( h_2 
    \left(\TmDel(z, t) - \TwDel(z, t)\right)\notag\\&\quad\qquad\qquad+h_3\left(\TaDel(z, t) - \TmDel(z, t)\right)\Big). \label{eqn:del_Tw_part_der}
\end{align}
\end{subequations}
Solving \eqref{eqn:del_Tm_part} for $\TwDel$
\begin{align}
    \TwDel(z,t) & = \frac{\vmDel(z,t)}{h_1}\pderiz\TmDel(z,t) + \TmDel(z,t) \label{eqn:del_Tw}
\end{align}
and substituting the resulting expression into \eqref{eqn:del_Tw_part_der} finally yields the desired \gls{dpde} 
(cf.\ \eqref{eqn:pde_subst:shifted:pde})
\begin{subequations}\label{eqn:filter_var}
\begin{equation}\label{eqn:filter_var_pde}
\begin{split}
    \vmDel(z,t) \pderitz\TmDel(z,t) & + h_1\pderit \TmDel(z,t) + \pderit\vmDel(z,t)\pderiz\TmDel(z,t) \\
                                                                         & + (h_2 + h_3)\vm(t)\pderiz\TmDel(z,t) + h_1h_3\frac{\vm(t)}{\vmDel(z,t)}\TmDel(z,t)\\
                                                                         & = 
                                                                         h_1h_3\frac{\vm(t)}{\vmDel(z,t)}\TaDel(z,t)
\end{split}
\end{equation}
with the boundary condition
\begin{equation}\label{eqn:filter_var_bc}
    \TmDel(0,t)  = \TminDel\left(t\right) = \Tmin(t - \tau(l,t)).
\end{equation}
\end{subequations}
For constant flow the \gls{dpde} model \eqref{eqn:filter_var} simplifies to
\begin{subequations}\label{eqn:pde_subst:shifted}
	\begin{multline}\label{eqn:pde_subst:shifted:pde}
	\pderitz\TmDel(z,t)+(h_2 + h_3)\pderiz\TmDel(z,t)  \\= 
	\frac{h_1}{\vm}\left(h_3(\TaDel(z,t)-\TmDel(z,t))-\pderit\TmDel(z,t)\right),
	\end{multline}
	with
	\begin{equation}
		\TmDel(0,t)=\Tmin(t-\tau(l)), \quad \Tm(l,t)=\TmDel(l,t).
	\end{equation}
\end{subequations}
Moreover, in the particular case $h_3=0$, i.e., for the perfectly isolated pipe, \eqref{eqn:filter_var} reduces to
\begin{multline*}
    \vmDel(z,t) \pderitz\TmDel(z,t) + h_1\pderit \TmDel(z,t) + 
    \pderit \vmDel(z, t)\pderiz\TmDel(z,t) \\
    + h_2 \vm(t) \pderiz\TmDel(z,t) = 0. 
\end{multline*}

\begin{figure}[!tb]
    \centering
    \begin{adjustbox}{width=0.5\textwidth}
        \input{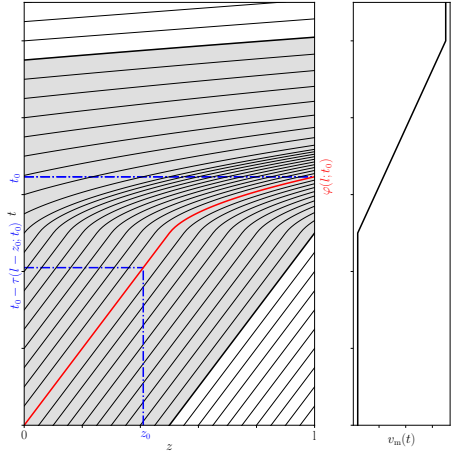}
    \end{adjustbox}
    \vspace{-3ex}\caption{Characteristics for a time variant flow regime.}
    \label{fig:filter_char}
\end{figure}

Remark, that the above performed computations correspond to the evaluation of the system \eqref{eqn:pde_1d} on the characteristic projections of 
\eqref{subeqn:pde_1d_Tm} (cf.\ 
\cite{Courant1962, John1991}). In this context, each of the functions
\begin{align*}
    [0,l]\ni z & \mapsto \varphi(z,t)\in\Re,\qquad t  \in\Re
 \end{align*}
simply corresponds to the particular characteristic projection $z\mapsto (z,\varphi(z,t))$ in the $(z,t)$-plane containing the point $(l,t)$. 
\figref{fig:filter_char} depicts the characteristic projections for a variable flow rate.
In the following section the approximation of the \gls{dpde}-model to a first order \gls{dde} is shown.

\section{Approximation as Delay-Differential Equation} \label{sec:approxDDE}
Within this section approximation schemes for the derived models are introduced. These approximations form the basis for the subsequent numerical
studies in Section \ref{sec:validation}. Moreover, the advantages and
disadvantages of the proposed models are discussed.

All \gls{pde}-models are semi-discretized w.r.t.\ the spatial variable only. This is achieved by means of the \gls{fdm}, i.e., by approximating
the spatial derivatives by backward differences: 
\begin{align*}
    \left(\pderiz \T\right)(z_i,t) &\approx \frac{\T[i](t) - \T[i-1](t)}{z_i - z_{i-1}},\quad i\in[1,n].
\end{align*}
Therein,  $n$ specifies the number of $n+1$ sampling points $z_0,\dots,z_n$. As 
a consequence the \gls{pde}-models are 
approximated by a system  consisting of $n$ \glspl{ode} respectively 
\glspl{dde}. For the sake of simplicity, these computations are discussed for constant flow rates only, i.e., for \eqref{eqn:pde_subst:shifted}.

The simplest approximation of the \gls{dpde}-model 
\eqref{eqn:pde_subst:shifted} 
with $n=1$  provides the link to the \gls{dde}-model 
\eqref{eqn:dde:approach}. It corresponds to a discretization of the spatial 
derivative by the difference
\begin{align*}
    \left(\pderiz \TmDel\right)\left(l,t\right) & \approx \frac{1}{l}\left(\TmDel\left(l,t\right)- \TmDel(0, t\right) = \frac{1}{l}\left(\Tm(t)
    - \Tmin(t - \tau(l))\right).
\end{align*}
Taking the input boundary condition into account,
\eqref{eqn:pde_subst:shifted} leads to
\begin{align}
    \begin{split}
        \pderit \Tm(t) & = k_1 \left(\Tmin(t - \tau(l)) - \Tm(t)\right) + k_2\left(\Ta(t) - \Tm(t)\right)\\
                       & \quad + k_3\pderit\Tmin(t - \tau(l)), \label{eqn:filter_dde}
    \end{split}
\end{align}
with the constants
\begin{align*}
    k_1 & = \frac{(h_2 + h_3) \vm}{\vm + h_1 l}, & k_2 & = \frac{h_1 h_3 l}{\vm + h_1 l}, & k_3 & = \frac{\vm}{\vm + h_1 l}.
\end{align*}
\Eqnref{eqn:filter_dde} reveals a similar structure as the heuristic 
\gls{dde}-model \eqref{eqn:dde:approach}.\footnote{Note that for the time 
variant case one reveals the time depending coefficients
	\begin{align*}
		k_1(t) & = \frac{\pderit \vm(t) + (h_2 + h_3) \vm(t)}{\vm(t) + h_1 l}, 
		& k_2(t) & = \frac{h_1 h_3 l}{\vm(t) + h_1 l}, & k_3(t) & = 
		\frac{\vm(t)}{\vm(t) + h_1 l},
\end{align*}
and the time variant delay $\tau(l, t)$.
} Apart from the different constants 
$k_i$ there is an additional term involving input temporal derivative, which is induced by a feedthrough of the input temperature in the
physical based approach. However, if the length of the pipe is sufficiently large
compared to the velocity of the medium or a sufficiently high heat exchange 
between medium and wall is present, this term can be neglected. In this case
\eqref{eqn:filter_dde} reduces to the \gls{dde}-model
\eqref{eqn:dde:approach}. Otherwise, the remaining term can be treated as described below for higher approximation orders.

Approximating the \gls{dpde}-model \eqref{eqn:pde_subst:shifted} by the backward 
difference leads to
\begin{align}
    \begin{split}\label{eqn:filter_dde_general}
        \derit \left( \TmDel[i](t) -  k_3 \TmDel[i-1](t)\right) & = k_1 \left(\TmDel[i-1](t) - \TmDel[i](t)\right) \\
                                                              & \quad + k_2\left(\TaDel[i](t) - \TmDel[i](t)\right),
    \end{split}
\end{align}
with the constants
\begin{align*}
    k_1 & = \frac{(h_2 + h_3) \vm}{\vm + h_1 \dz}, & k_2 & = \frac{h_1 h_3 \dz}{\vm + h_1 \dz}, & k_3 & = \frac{\vm}{\vm + h_1 \dz},
\end{align*}
including the constant spatial step $\dz = z_i - z_{i-1}$.
\Eqnref{eqn:filter_dde_general} results in a system of \glspl{dde} of the form
\begin{align*}
	\derit \bTm(t) &= \bAm \, \bTm(t) + \bbm[1] \derit \Tmin(t) + \bbm[2] 
	\Tmin(t) + \bDm \bTa(t),
\end{align*}
with the corresponding system matrix $\bAm \in \Re^{n \times n}$, input vectors 
$\bbm[1],\ \bbm[2] \in \Re^{n}$ and disturbance matrix $\bDm \in \Re^{n \times n}$. The state $\bT$ can be noted with
\begin{align*}
    \bT(t) & = \begin{pmatrix}
        \TDel[1](t), & \TDel[2](t), & \ldots, & \TDel[n](t)
        \end{pmatrix}^T.
\end{align*}
In order to eliminate the time derivative of the inflow temperature the transformation
\begin{align*}
	\bTmh(t) &= \bTm(t) - \bbm[1] \Tmin(t)
\end{align*} is applied, which reveals the state space description
\begin{align*}
	\derit \bTmh(t) &= \bAm \, \bTmh(t) + \left(\bAm \bbm[1] + \bbm[2] \right)
	\Tmin(t) + \bDm \bTa(t).
\end{align*}

In case of variable flow the previously described steps can be applied in a similar way. However, the system must be additionally discretized 
with respect to time, due to the time dependent slope of the characteristics.

\section{Simulation and Experimental Validation}\label{sec:validation}
In this section the different modeling approaches are compared in simulation 
studies and validated with measurement data. All further analyses consider the medium to be water. The discussed 
models are the one-dimensional \gls{pde}-model \eqref{eqn:pde_1d} approximated by the \gls{fdm} with a high resolution 
of $201$ discretization points ($n=200$), the proposed \gls{dpde}-approach 
\eqref{eqn:filter_var} with a low-order \gls{fdm} approximation with $6$ sampling points ($n=5$), named \acrshort{dpde5}, the 
\gls{dde}-model 
\eqref{eqn:dde:approach}, 
the adapted \gls{dde}-model \eqref{eqn:adapted_common_dde}, and the \acrshort{dpde1}-model 
\eqref{eqn:filter_dde} derived from the proposed \gls{dpde}-model with $2$ discretization points ($n=1$).

For a numerical comparison of the simulations and measurements the \gls{rms} error
\begin{align*}
    \error[2](z) & = \sqrt{\frac{1}{p}\sum_{j=1}^{p}\left|\T^j(z) - \tildeT^j(z)\right|^2}
\end{align*}
and the maximum error metric
\begin{align*}
    \error[\infty](z) & = \max_{1<j<p}\left|\T^j(z) - \tildeT^j(z)\right|
\end{align*}
are introduced. Therein $\T$ and $\tildeT$ denote the benchmark and simulation data at a specific spatial position $z$ over all times $j \in
\left[t_1,\; t_p\right]$, respectively.

The simulation study, identification, and validation is performed in Python.

\subsection{Simulation study}
The simulation study captures a scenario with a temperature ramp from \SI{20}{\celsius} to \SI{60}{\celsius}
of the pipe input temperature $\Tmin$ to unveil the main differences between the five approaches. Therefore, the high-order \gls{pde}-model is  
defined as benchmark.
For the simulation a $l=\SI{5}{\meter}$ long stainless steel pipe with an inner radius $\Rm = \SI{7,7}{\milli\meter}$ and an 
outer radius $\Rw = \SI{10,65}{\milli\meter}$ is considered. Moreover, a constant 
medium velocity of $\vm=\SI{0.5}{\meter\per\second}$ is assumed. 
\tabref{tab:params} provides an overview of the physical parameters, 
which are chosen according to the common literature \cite{Rohsenow1998}.
The heat transfer coefficient $\alphama$ of the \gls{dde}-model 
\eqref{eqn:adapted_common_dde} is calculated by means of \eqref{eqn:alpha_ma}. Afterwards, the correction factor $\epsilon$ is determined by means of a 
least squares optimization based on simulation results\footnote{Note that a different scenario, where the input temperatures is 
decreased from \SIrange{80}{30}{\celsius} at a medium velocity of \SI{0.4}{\meter\per\second} is used for the identification.} of the benchmark 
model (cf.\ \tabref{tab:params}).

\begin{table}
    \setlength{\extrarowheight}{0.25em}
	\centering 
	\caption{Used parameters for validation studies.\label{tab:params}}
	\begin{tabular}{|c|c|c|c|}
		\hline 
		\textbf{Parameter} & \textbf{Simulation} & \textbf{Measurement} & 
		\textbf{Unit} \\ \hline
		\rhow & \multicolumn{2}{c|}{\SI{7856} } & 
		\si{\kilogram\per\cubic\meter} \\[0.25em] \hline 
		\cpw & \multicolumn{2}{c|}{\SI{500}} & 
		\si{\joule\per\kilo\gram\per\kelvin} 
		\\[0.25em] \hline 
		\lambdaw & \multicolumn{2}{c|}{\SI{20}}  & 
		\si{\watt\per\meter\per\kelvin} \\[0.25em] \hline 
		\rhom &\multicolumn{2}{c|}{\SI{997.04}} & 
		\si{\kilogram\per\cubic\meter} \\[0.25em] \hline 
		\cpm &\multicolumn{2}{c|}{\SI{4179}}  & 
		\si{\joule\per\kilo\gram\per\kelvin} 
		\\[0.25em] \hline  
		l & \SI{5} & \SI{1.62} & \si{\meter} \\ \hline
		$\epsilon$ & \SI{0.7} & \SI{0.91} & -\\[0.25em]
        \hline
		\alphamw & \SI{1000} & \SI{3052.87} & 
		\si{\watt\per\square\meter\per\kelvin} 
        \\[0.25em] \hline
        \alphawa & \SI{80} & \SI{46.98} & 
        \si{\watt\per\square\meter\per\kelvin} \\[0.25em]
		\hline
        \alphama & \SI{73.39} & \SI{46} & 
        \si{\watt\per\square\meter\per\kelvin} \\[0.25em]
		\hline
	\end{tabular}
\end{table}

\figref{fig:compare_ramp_pde_filter_dde} presents the results for a ramp input temperature $\Tmin$. It can be observed that the low order 
\acrshort{dpde5}-model reveals nearly the same results as the high-order \gls{pde}-model. Especially no numerical diffusion effects 
\cite{Andersson2011} due to 
the \gls{fdm} approximation can be observed for the \acrshort{dpde5}-model. Moreover, an good coincidence can be also observed for the 
\acrshort{dpde1}. 
In contrast, the \gls{dde}-model and adapted \gls{dde}-model reveal a twenty respectively a ten times higher \gls{rms} 
error than the 
\acrshort{dpde5}-model. 
Compared with the \acrshort{dpde1}-model still an approximately four and two times higher \gls{rms} error (cf.\ 
\tabref{tab:errorSimStud}) can be observed. 
Moreover, \tabref{tab:errorSimStud} reveals that the maximum error of the adapted \gls{dde}-model is $2.5$ times higher than the 
maximum error 
of the \acrshort{dpde1}.
Furthermore, the stationary inaccuracy of the \gls{dde}-model can be reduced using the adapted \gls{dde}-model with
adapted parameters.

\begin{figure}
    \centering
    \begin{adjustbox}{width=\textwidth}
        \input{images/compare_ramp_pde_filter_dde.tex}
    \end{adjustbox}\vspace*{-9mm}
    \caption{Medium temperature comparison of the new \acrshort{dpde5}-model, the \gls{pde}-model, and 
    the \gls{dde}-approaches with a ramp for $\Tmin$ at $z=l$.}
    \label{fig:compare_ramp_pde_filter_dde}
\end{figure}

\begin{figure}
	\centering
    \begin{adjustbox}{width=\textwidth}
		\input{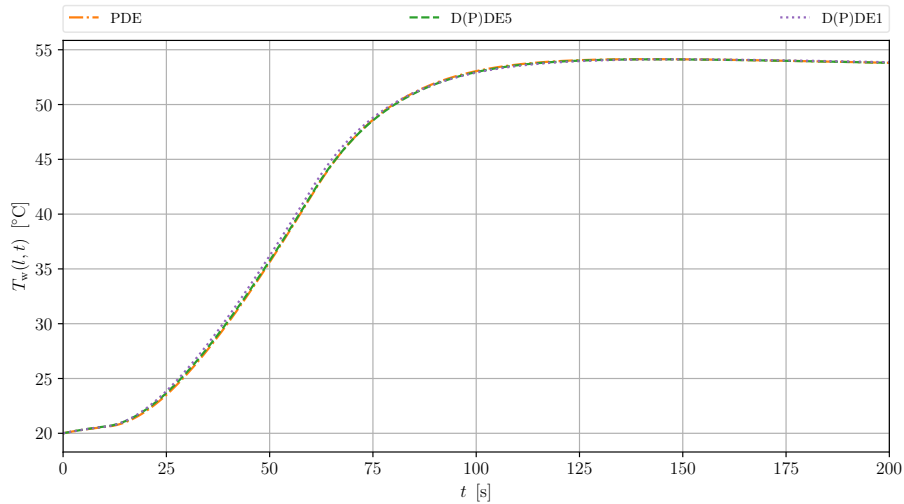}
	\end{adjustbox}\vspace*{-9mm}
	\caption{Wall temperature comparison of the new \acrshort{dpde5}-model, the \gls{pde}-model, and 
    the \gls{dde}-approaches with a ramp for $\Tmin$ at $z=l$.}
	\label{fig:compare_ramp_wall_pde_filter}
\end{figure}

Another advantage of the new \gls{dpde}-model is that the wall temperature can be 
reconstructed easily by means of \eqref{eqn:del_Tw}. A comparison of the wall temperatures of the high-order 
\gls{pde}-model and the \gls{dpde}-models is presented in
\figref{fig:compare_ramp_wall_pde_filter}.

\begin{table} \label{tab:errorSimStud}
\caption{Error measures of \acrshort{dpde5}-, \acrshort{dpde1}-, adapted \gls{dde}- and \gls{dde}-model.}
    \centering
    \begin{tabular}{llccccc}\hline\hline
               &       & \gls{dde} & adapted \gls{dde} & \acrshort{dpde1} & \acrshort{dpde5} & Units \\\hline\hline
        medium & $\error[2](l)$        & $1.2483$ & $0.6435$ & $0.3087$ & $0.0619$ & \si{\celsius} \\
               & $\error[\infty](l)$   & $2.5454$ & $1.6612$ & $0.7077$ & $0.1791$ & \si{\celsius} \\\hline
        wall   & $\error[2](l)$        & - & - & $0.2672$ & $0.0849$ & \si{\celsius} \\
               & $\error[\infty](l)$   & - & - & $0.5907$ & $0.1823$ & \si{\celsius} \\\hline\hline
    \end{tabular}
\end{table}

\subsection{Measurement}
\begin{figure}
	\centering
	\includegraphics[width=\textwidth]{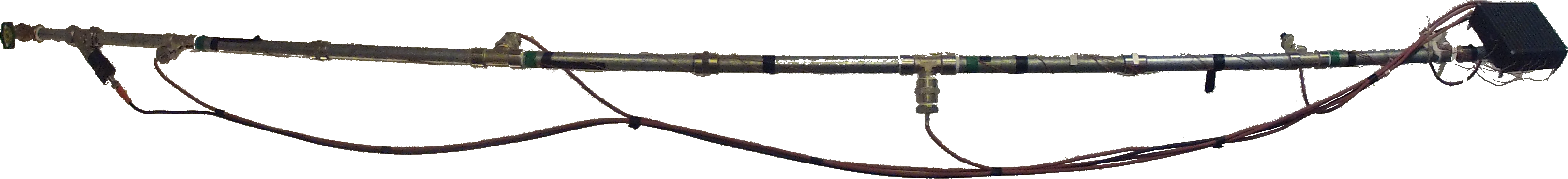}
	\vspace*{-8mm}
	\caption{Pipe test rig used for validation.}
	\label{fig:testrig}
\end{figure}

\begin{figure}
    \centering
    \begin{adjustbox}{width=\textwidth}
        \input{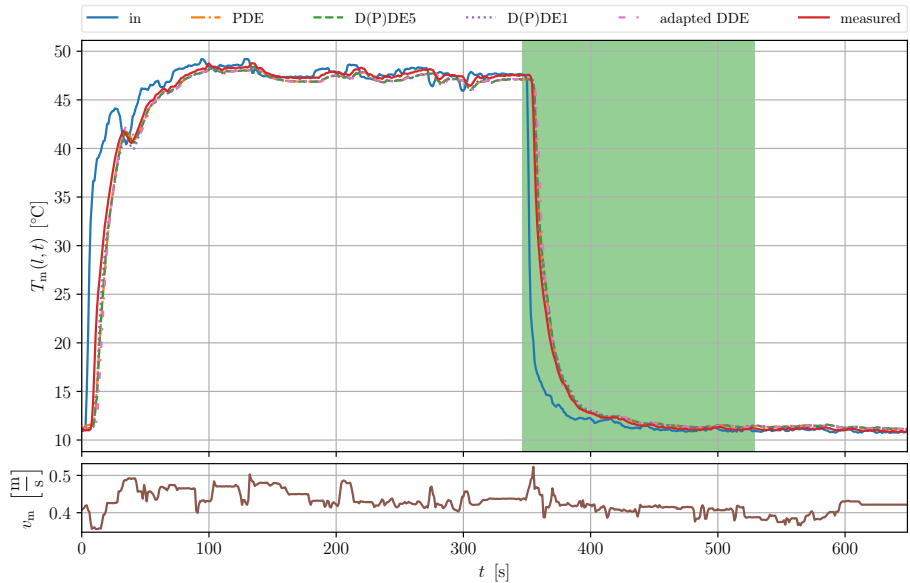}
    \end{adjustbox}\vspace*{-9mm}
    \caption{\markChange{Measurement with green marked data, that is used for the identification of the parameters $\alphamw$, $\alphawa$, $\alphama$
    and $\epsilon$.}}
    \label{fig:ident_meas_pde_filter_dde}
\end{figure}

\begin{figure}
    \centering
    \begin{adjustbox}{width=\textwidth}
        \input{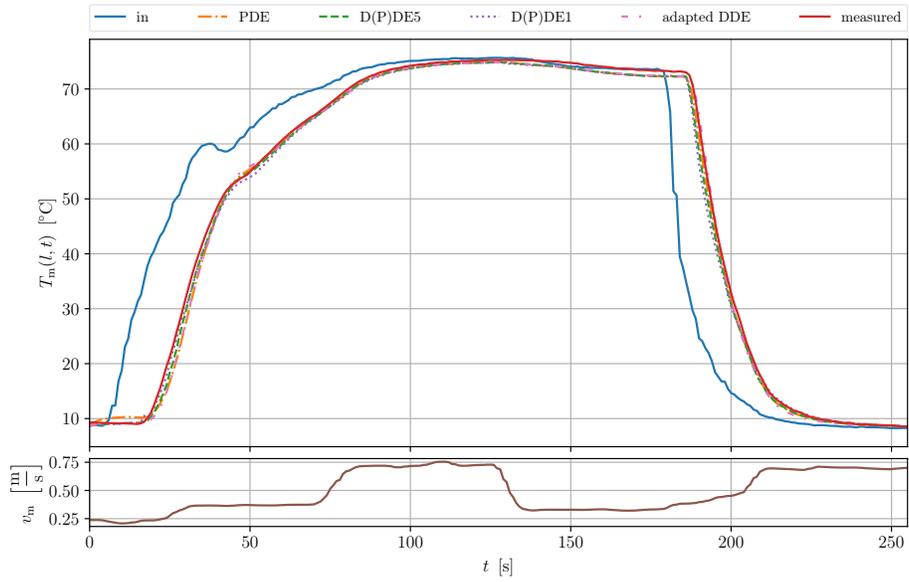}
    \end{adjustbox}\vspace*{-9mm}
    \caption{Medium temperature comparison of the new \acrshort{dpde5}-model, the \gls{pde}-model, and 
    the \gls{dde}-approaches with measurement data at $z=l$.}
    \label{fig:compare_meas_pde_filter_dde}
\end{figure}

\begin{figure}
    \centering
    \begin{adjustbox}{width=\textwidth}
        \input{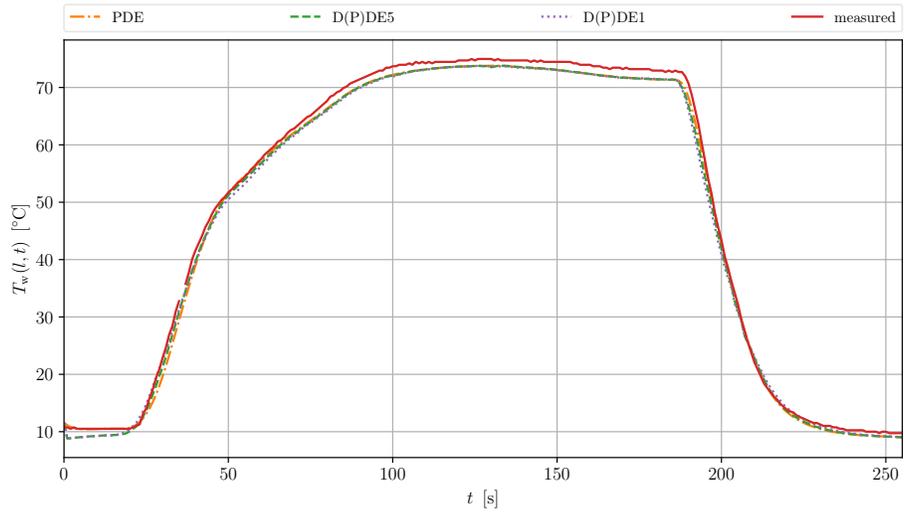}
    \end{adjustbox}\vspace*{-9mm}
    \caption{Wall temperature comparison of the new \acrshort{dpde5}-model and the \gls{pde}-model with measurement data at $z=l$.}
    \label{fig:compare_meas_wall_pde_filter}
\end{figure}

In this subsection the one-dimensional \gls{pde}-model \eqref{eqn:pde_1d}, the 
new \acrshort{dpde5}-model \eqref{eqn:filter_var}, the adapted \gls{dde}-model, and the \acrshort{dpde1}-model 
\eqref{eqn:filter_dde} are experimentally validated. 
This was achieved by means of the test rig depicted in \figref{fig:testrig}, which was specifically designed for the validation of the analyzed
modeling approaches. The pipe is filled with water. At each \SI{0.54}{\meter} both the medium temperature and the wall 
temperature are measured (cf. \figref{fig:general_pipe}). Thus, with a total 
length of \SI{1,62}{\meter}, four measuring points are available, the 
first of which is used as input temperature. Hence, three 
points are left for the validation. The medium and wall temperatures are 
measured by PT100 sensors and thermocouples, respectively. Furthermore, the 
volume flow rate through  the pipe is measured by applying the principle of 
differential pressure and the ambient temperature is measured by another 
thermocouple. The signal processing is done by an Arduino Uno complemented with  appropriate 
sensor boards. Due to its slow variation the ambient temperature is set to 
a constant value of \SI{25.8}{\celsius}. Moreover, the dynamics of the PT100 
sensors are compensated by means of an inverse model. The pipe under consideration has an inner diameter of 
\SI[fraction-function=\sfrac]{3/5}{inch} and an outer 
diameter of \SI[fraction-function=\sfrac]{4/5}{inch}. 
\figref{fig:compare_meas_pde_filter_dde}~-~\ref{fig:compare_meas_intermediate_pde_filter} and \tabref{tab:numerical_results_measurements}
present the validation results, where the input temperature and the volume flow 
rate are varied.
The required heat transfer coefficients $\alphamw$, $\alphawa$, and the correction parameter $\epsilon$ are
determined by a least squares optimization based on the measured medium and wall temperatures  at the pipe outlet. Afterwards, the heat transfer
coefficient $\alphama$ is computed by evaluating \eqref{eqn:alpha_ma} with the identified values of $\alphamw$ and $\alphawa$.
For this purpose, another 
data set is used (cf.\ highlighted sector in \figref{fig:ident_meas_pde_filter_dde}), where the velocity of the medium is nearly constant. The 
identified and computed parameters are given in 
\tabref{tab:params}.
\figref{fig:compare_meas_pde_filter_dde} presents the measured input and output 
temperature, the medium velocity, and the simulated output temperatures. An
almost perfect match of all models with the measured output can be observed. Based on the error measures presented in 
\tabref{tab:numerical_results_measurements} one can observe that the \acrshort{dpde5}-model reveals the smallest average error.
However, if just the medium temperature at the output of the pipe is required each of the models is applicable. In contrast, if the wall temperature or 
an 
intermediate medium temperature is needed the \acrshort{dpde5}-model or its simplest solution, the \acrshort{dpde1}, are a good
alternatives to the 
\gls{pde}-model.\footnote{Using \eqref{eqn:filter_dde} or \eqref{eqn:adapted_common_dde} for a temperature calculation 
at a point $z_0 < l$ of the pipe, the length $l$ has to be replaced by the chosen point $z_0$. Thus, the shell surface area in 
\eqref{eqn:adapted_common_dde} is changing too.}
Furthermore, at $z = \nicefrac{l}{3}$ all 
the newly proposed models show a perfect match with the measured data (cf.\ 
\figref{fig:compare_meas_intermediate_pde_filter}), whereas the error of the adapted \gls{dde}-model increases. A spatially dependent definition (resp.\ identification) of the 
correction factor $\epsilon$ may lead to better results.
Moreover, the wall 
temperatures calculated by the \gls{pde}- and the \acrshort{dpde5}-model show a similar 
behavior as the measurement. The occurring offset of \SI{2}{\celsius} (compare \figref{fig:compare_meas_wall_pde_filter} between 
\SIrange{100}{180}{\second}) is likely to be
caused by the nonlinear behavior of the thermocouple, which is not compensated. At different measuring points different offsets (positive and 
negative) arise. 

To sum up, the proposed modeling 
approaches succeed in reproducing the measurements in  the considered scenario. 
Furthermore, in contrast to the classical \gls{dde}-approach, the \gls{pde}-model 
and the \gls{dpde}-approaches admit for the additional computation of the wall 
temperature.

\begin{figure}
	\centering
    \begin{adjustbox}{width=\textwidth}
		\input{images/compare_meas_intermediate_pde_filter.tex}
	\end{adjustbox}\vspace*{-9mm}
    \caption{Medium temperature comparison of the new \acrshort{dpde5}-model, the \gls{pde}-model, and the \gls{dde}-models with measurement data 
    at $z=\nicefrac{l}{3}$.}
	\label{fig:compare_meas_intermediate_pde_filter}
\end{figure}

\begin{table}
    \caption{Numerical comparison of adapted \gls{dde}-, \acrshort{dpde5}-, \acrshort{dpde1}- and \gls{pde}-model against 
    measurement data.}
    \label{tab:numerical_results_measurements}
    \centering
    \begin{tabular}{llccccc}\hline\hline
               &                       & adapted \gls{dde}  & \acrshort{dpde1} & \acrshort{dpde5} & \gls{pde} & Unit\\\hline\hline
        medium & $\error[2](l)$        & $1.2605$ & $1.4289$  & $1.1913$ & $1.2876$ & \si{\celsius}\\
               & $\error[\infty](l)$   & $4.7591$ & $7.3504$  & $5.0512$ & $5.0342$ & \si{\celsius}\\
               & $\error[2]\left(\nicefrac{l}{3}\right)$      & $1.2869$ & $0.4921$ & $0.8212$ & $0.6690$ & \si{\celsius}\\
               & $\error[\infty]\left(\nicefrac{l}{3}\right)$ & $6.8591$ & $2.0078$ & $3.9308$ & $2.9872$ & \si{\celsius}\\\hline
        wall   & $\error[2](l)$        & - & $2.5682$ & $2.4378$ & $2.3330$ & \si{\celsius}\\
               & $\error[\infty](l)$   & - & $33.0015$ & $32.7530$ & $31.3974$ & \si{\celsius}\\\hline\hline
    \end{tabular}
\end{table}

\section{Conclusion and Further Work}

\subsection{Conclusion}
This contribution presents a physical derivation of commonly used 
\gls{dde}-models describing the thermal behavior of a plug 
flow through a pipe, where a heat transfer between the transport medium and the 
wall and between the wall and the ambient is considered explicitly. Starting from a 
one-dimensional \gls{pde}-model for the fluid inside the pipe and a two-dimensional \gls{pde}-model for the wall of the pipe a one-dimensional
\gls{pde}-model is derived. Based on the latter one a novel \gls{dpde}-model is introduced, which is a combination of the \gls{pde}- and
\gls{dde}-approach. On the basis of the
\gls{dpde}-model a new version of the well known \gls{dde}-model is derived. It is 
shown that the common heuristic approach has to be extended by a tuning 
parameter to be able to cover the pipe dynamics. In contrast no tuning 
parameter has to be applied to the proposed \gls{dpde}-model and the new 
\gls{dde}-approach. Moreover, the new \gls{dpde}-model allows to calculate the wall 
temperature at any position. The different models are compared in simulation to 
illustrate their strengths and weaknesses. It has been shown that for highly 
transient changes at the input the new modeling approach delivers better 
results than the common \gls{dde}-approach. Compared with a \gls{fdm} 
simulation of the one-dimensional \gls{pde} no numerical diffusion effect can be observed 
for the new approach. Finally, a validation against measurements shows an almost perfect 
match of all modeling approaches.

\subsection{Further work}
In a first step the introduced \gls{dpde}-model can be used to observe the temperature 
profile in plug flow tube reactors (e.g.\ in a catalyst). The obtained \gls{dpde}-model 
forms the basis for an input-output description for temperature of the medium 
involving measured boundary quantities only. The model forms an appropriate basis for
the identification of the heat transfer coefficients, if the latter are unknown. In contrast
to alternative schemes, which require numerically expensive optimization an comparably simple 
approach \cite{Knppel2013ACT} can be used, which requires only very basic 
optimization algorithms.
Furthermore, the observed data can be used to govern such profiles by appropriate control 
algorithms. In addition the applicability of the \gls{dpde}-model to pipe networks will be 
investigated in the future. Finally, in view of applications with non-turbulent flow regimes the a priori spatially one-dimensional 
modeling approaches are likely to be not sufficiently accurate. This motivates further investigations on the basis of higher dimensional 
stationary-flow regimes as studied numerically for example in \cite{Cao2019}.

\section{Acknowledgments}
The present contribution is a result of the research project MoReNe (FFG-Nr. 
864725) funded by the Austrian Research Promotion Agency (FFG) and Innio 
Jenbacher GmBH \& Co OG located in Jenbach (Austria). The authors gratefully 
acknowledge Jonathan Halmen, who has set up the long pipe test rig and recorded 
the measurements.

\appendix

\section{Approximate heat transfer coefficients}\label{app:overallcoefficients}
Within this section the calculation of the overall heat transfer coefficients 
for the defined mean temperature \eqref{eqn:pde_wall_1d_mean} is explained.
The stationary solution of \eqref{eqn:pde_wall_3d} satisfies
\begin{equation*}
	\pderir\left( r \dq(r,z) \right) = 0,\quad \dq(r,z)=-\lambdaw \pderir  \Tw(r,z).
\end{equation*}
Integrating the first of these equations w.r.t.\ $r$ over the interval $[R_m,r]$ leads to
\begin{equation*}
  r \dq(r,z) - \Rm \dq(\Rm,z) = 0.
\end{equation*}
With $\dq(r,z)=-\lambdaw \pderir\Tw(r,z)$ one obtains
\begin{equation*}
\lambdaw r \pderir \Tw(r,z) + \Rm \dq(\Rm,z) = 0.
\end{equation*}
Solving this \gls{ode} for $r\mapsto \Tw(r,z)$ yields
\begin{equation*}
	\lambdaw \left(\Tw(r, z) - \Tw(\Rm, z)\right) = \ln\left( \frac{\Rm}{r} 
	\right) \Rm \dq(\Rm, t).
\end{equation*}
The latter equation is integrated over $\QAw$ 
\begin{align*}
	 \int_{\Rm}^{\Rw} 2\pi r \lambdaw \left(\Tw(r,z) - \Tw(\Rm, z)\right)\dd{r} &= 2 \pi \Rm \dq(\Rm, 
	t) \int_{\Rm}^{\Rw} r \ln\left( \frac{\Rm}{r} \right) \dd{r} 
  \end{align*}
to obtain
  \begin{multline*}
  \Aw \lambdaw \left(\barTw(z) - \Tw(\Rm, z)\right)\\                                                          = 2 \pi \Rm  \left(\Rw^2 \ln\left(\frac{\Rw}{\Rm}\right) + \frac12\left(\Rm^2 - 
	\Rw^2\right)\right)\,\dq(\Rm, 
	t).
\end{multline*}
The latter equation can be simplified by substituting $\Aw = \pi (\Rw^2 - \Rm^2)$ and eliminating
$\Tw(\Rm,z)$ by means of the boundary condition \eqref{subeqn:bc_mw}. This finally yields
\begin{align}
	\Tm(z) - \barTw(z)  &=  \left(\frac1\alphamw + 
	\frac{\barRm}{\lambdaw}\right)\dq(\Rm,z)
	\label{eqn:stationary_sol_meanTw_Tw}
\end{align}
with
\begin{align*}
	\barRm &= 
	\Rm\left(\frac{\Rw^2}{\Rw^2-\Rm^2}\ln\left(\frac{\Rw}{\Rm}\right) - 
	\frac12\right).
\end{align*}
Similar computations for the outer boundary of the jacket lead to
\begin{align}
	\barTw(z) - \Ta &=  \left(\frac1\alphawa + 
	\frac{\barRw}{\lambdaw}\right)\,\dq(\Rm,z),
	\label{eqn:stationary_sol_meanTw_Ta}
\end{align}
with
\begin{align*}
\barRw &= \Rm \left(-\frac{\Rm^2}{\Rw^2 - \Rm^2} 
\ln\left(\frac{\Rw}{\Rm}\right) + \frac12\right).
\end{align*}
Summing up \eqref{eqn:stationary_sol_meanTw_Tw} and 
\eqref{eqn:stationary_sol_meanTw_Ta} reveals
\begin{align*}
 \Tm(z) - \Ta = \dq(\Rm, z)\left(\frac1\alphamw + \frac1\alphawa + 
 \frac{\Rm}{\lambdaw} \ln\left(\frac{\Rw}{\Rm}\right) \right)
\end{align*}
the well known formulation of the overall heat transfer for a cylindrical pipe
(cf.\ \cite[p.~31~ff.]{Baehr2011}).

\bibliography{bib/sources.bib}

\end{document}